\documentclass[times,sort&compress,final]{elsarticle}

\usepackage{hyperref}
\hypersetup{
    colorlinks=true,
    linkcolor=blue,
    filecolor=magenta,      
    urlcolor=cyan,
}
\usepackage{dirtree}
\usepackage{cpc}
\usepackage{framed,multirow}
\usepackage{nameref}
\usepackage[labelfont=bf,justification=justified,singlelinecheck=true]{caption}
\usepackage{amsmath}
\usepackage{bm}
\usepackage{soul}
\usepackage{amssymb}
\usepackage{commath}
\usepackage{latexsym}
\usepackage{braket}
\usepackage{url}
\usepackage[table]{xcolor}
\usepackage{listings}
\usepackage{verbatimbox}
\usepackage{tikz}
\usepackage[english]{babel}
\usepackage{blindtext}
\usepackage[symbol]{footmisc}

\DeclareMathOperator{\arcsinh}{arcsinh}

\definecolor{codegreen}{rgb}{0,0.6,0}
\definecolor{codegray}{rgb}{0.5,0.5,0.5}
\definecolor{codepurple}{rgb}{0.58,0,0.82}
\definecolor{backcolour}{rgb}{0.97,0.97,0.97}

\lstdefinestyle{myCPPstyle}{
    backgroundcolor=\color{backcolour},   
    commentstyle=\color{codegreen},
    keywordstyle=\color{magenta},
    numberstyle=\tiny\color{codegray},
    stringstyle=\color{codepurple},
    basicstyle=\ttfamily\footnotesize,
    breakatwhitespace=false,         
    breaklines=true,                 
    captionpos=b,                    
    keepspaces=true,                                  
    numbersep=5pt,                  
    showspaces=false,                
    showstringspaces=false,
    showtabs=false,                  
    tabsize=2,
    language=C++,
    numbers=none, 
    xleftmargin=5mm, 
    xrightmargin=5mm,
    frame=single,
    framerule=0pt
}
\lstdefinestyle{myBASHstyle}{
    backgroundcolor=\color{backcolour},   
    commentstyle=\color{codegreen},
    keywordstyle=\color{magenta},
    numberstyle=\tiny\color{codegray},
    stringstyle=\color{codepurple},
    basicstyle=\ttfamily\footnotesize,
    breakatwhitespace=false,         
    breaklines=true,                 
    captionpos=b,                    
    keepspaces=true,                                  
    numbersep=5pt,                  
    showspaces=false,                
    showstringspaces=false,
    showtabs=false,                  
    tabsize=2,
    language=bash,
    numbers=none, 
    xleftmargin=5mm, 
    xrightmargin=5mm
}

\definecolor{light-gray}{gray}{0.97}
\newcommand{\listingbox}[1]{%
         \begin{center}%
            \begin{tikzpicture}%
                \node[rectangle, draw=gray, fill=light-gray, inner xsep=5pt, inner ysep=6pt, outer ysep=10pt]{
                \begin{minipage}{0.85\linewidth}#1\end{minipage}};%
            \end{tikzpicture}%
         \end{center}%
}

\def\qtraj{{\tt QTraj} }
\def\ba{\begin{eqnarray}}
\def\ea{\end{eqnarray}}
\def\be{\begin{equation}}
\def\ee{\end{equation}}
\def\als{\alpha_s}

\newcommand\preprintnumber[1]{%
    \bgroup
    \renewcommand{\thefootnote}{\fnsymbol{footnote}}
    \footnote[0]{Preprint numbers:  #1}
    \egroup
}

\journal{Computer Physics Communications}

\begin{document}

\verso{Ba Omar \textit{etal}}

\begin{frontmatter}

\title{QTRAJ 1.0:  A Lindblad equation solver for heavy-quarkonium dynamics}

\author[1]{Hisham \snm{Ba Omar}}
\author[2]{Miguel \'{A}ngel \snm{Escobedo}}
\author[1]{Ajaharul \snm{Islam}}
\author[1]{Michael \snm{Strickland}\corref{cor1}
\cortext[cor1]{Corresponding author: mstrick6@kent.edu }}
\author[1]{Sabin \snm{Thapa}}
\author[3]{Peter \snm{Vander Griend}}
\author[4]{Johannes Heinrich \snm{Weber}}

\address[1]{Department of Physics, Kent State University, Kent, OH 44242, United States}
\address[2]{Instituto Galego de F\'{i}sica de Altas Enerx\'{i}as (IGFAE), Universidade de Santiago de Compostela. E-15782, Galicia, Spain}
\address[3]{Physik-Department, Technische Universit\"{a}t M\"{u}nchen, James-Franck-Str. 1, 85748 Garching,
Germany}
\address[4]{Institut f\"ur Physik, Humboldt-Universit\"at zu Berlin \& IRIS Adlershof, D-12489 Berlin, Germany}

\availableonline{\today}
\communicated{M. Strickland}

\begin{abstract}
We introduce an open-source package called \qtraj that solves the Lindblad equation for heavy-quarkonium dynamics using the quantum trajectories algorithm.  
The package allows users to simulate the suppression of heavy-quarkonium states using externally-supplied input from 3+1D hydrodynamics simulations.
The code uses a split-step pseudo-spectral method for updating the wave-function between jumps, which is implemented using the open-source multi-threaded FFTW3 package.  
This allows one to have manifestly unitary evolution when using real-valued potentials.  
In this paper, we provide detailed documentation of \qtraj 1.0, installation instructions, and present various tests and benchmarks of the code.

\vspace{5mm}

\begin{small}

\noindent 
{\bf Program Summary}

\vspace{3mm}

\noindent
{\em Program Title:}  QTraj 1.0 \\
{\em Developer's repository link:} https://bitbucket.org/kentphysics/qtraj-fftw \\
{\em CPC Library link to program files: } \\
{\em Licensing provisions(please choose one):} GPLv3 \\
{\em Programming language:} C  \\
{\em Nature of problem:}\\
Solution of the Lindblad equation for evolution of the heavy-quarkonium reduced density matrix in a quark-gluon plasma.\\
{\em Solution method:}\\
We use the quantum trajectories algorithm to solve the Lindblad equation.  \\
\end{small}

\end{abstract}


\preprintnumber{TUM-EFT 142/21; HU-EP-21/17-RTG}

\end{frontmatter}


\section{Introduction}
\label{sec:intro}

Quantum Chromodynamics (QCD) is the quantum field theory that describes the strong nuclear force. 
The degrees of freedom contained in the QCD Lagrangian are quarks and gluons, which are collectively called partons.  
At low temperatures ($T \lesssim 2 \times 10^{12}$ K), partons are confined inside hadrons such as protons and neutrons, and, as a result, they are never observed in isolation. 
At temperatures higher than this threshold or, alternatively, at sufficiently high baryon density, QCD predicts a phase transition to a deconfined quark-gluon plasma (QGP) \cite{Bazavov:2013txa,Borsanyi:2016bzg,Bazavov:2020teh}. 
The QGP can be studied on Earth by colliding ultrarelativistic heavy ions, such as lead or gold, with ongoing experiments being performed at both the Relativistic Heavy Ion Collider at Brookhaven National Laboratory and the Large Hadron Collider at the European Organization
for Nuclear Research (CERN)~\cite{Averbeck2015,Busza:2018}. 

In practice, it is very challenging to extract dynamical information about the QGP from experimental data since this new state of matter exists for only a very short time after the collision, $t_\text{QGP} \lesssim 15\ \text{fm/c}$. 
It was theorized in \cite{Matsui:1986dk,Karsch:1987pv} that heavy quarkonium, a colorless bound state consisting of a heavy quark and a heavy anti-quark, melts inside of the QGP and, therefore, quarkonium suppression in heavy-ion collisions could be used as an important signature of the formation of a QGP. 
The theoretical description of in-medium quarkonium dynamics has changed significantly since these early studies. 
In the beginning, it was thought that the main phenomenon responsible for quarkonium suppression was color screening. Later, however, other phenomena were identified, which are equally or more important \cite{Laine:2006ns,Strickland:2011mw,Rothkopf:2019ipj,Akamatsu:2020ypb,Rothkopf:2020vfz}. 
In particular, one finds that collisions between heavy-quark bound states and particles in the medium introduce dispersion, which results in an imaginary contribution to the heavy-quark potential \cite{Laine:2006ns,Brambilla:2008cx,Brambilla:2011sg,Brambilla:2013dpa}. 
In fact, in the small coupling limit, collisions dissociate quarkonium bound states already at small distances where screening is not yet relevant. 
In addition to collisional dissociation, one must also consider recombination \cite{BraunMunzinger:2000px,Thews:2000rj}, a process in which a heavy quark and anti-quark, which could be initially uncorrelated, form a new bound state inside the medium or during late-time hadronic freeze-out.  
Statistical recombination is potentially very important for charm quarks, as opposed to bottom quarks, since charm quarks are more copiously produced in heavy-ion collisions~\cite{Emerick:2011xu}; however, local quantum regeneration could also play a role.

As a result, we need a theoretical description that can self-consistently include all three of these mechanisms.
In recent years, there have been many advances in the application of the open quantum system formalism to heavy quarkonium dynamics \cite{Borghini:2011ms,Akamatsu:2012vt,Akamatsu:2014qsa,Blaizot:2015hya,Katz:2015qja,Brambilla:2016wgg,Brambilla:2017zei,Kajimoto:2017rel,Blaizot:2017ypk,Akamatsu:2018xim,Yao:2018nmy,Yao:2018sgn,Blaizot:2018oev,Yao:2020eqy,Yao:2021lus}. 
In this approach, information about the quantum state of quarkonium is encoded in a reduced density matrix and the equation that describes the evolution of the reduced density matrix is called a master equation. 
In this paper, we focus on the master equations developed in \cite{Brambilla:2016wgg,Brambilla:2017zei}. 
Therein, it was found that, if the temperature $T$ is much larger than the binding energy $E$, the master equation is Markovian. 
It is a well-known result in the field of open quantum systems that all Markovian evolution equations that conserve the properties of the density matrix (trace-preserving, etc.) are Lindblad equations \cite{Gorini:1975nb,Lindblad:1975ef}. 
In \cite{Brambilla:2016wgg,Brambilla:2017zei}, the Lindblad equation was solved numerically in order to obtain phenomenological results. 
This was done in the following way:
\begin{itemize}
\item Expanding the density matrix in spherical harmonics and truncating at a given maximum angular momentum. In this way, the dimensionality of the problem is reduced from three to one dimensions.
\item Discretizing the one dimensional space on a finite lattice. 
\item Using the \textit{mesolve} function of the publicly available package QuTiP 2 \cite{qutip1,qutip2}.
\end{itemize}

However, as the size of the angular momentum basis and one-dimensional lattice used to solve the Lindblad equation become large, a huge amount of computational resources are required. 
For example, a naive counting indicates that doubling the size of the lattice increases the computational cost by a factor of at least four. 
This is a well-known problem faced when solving the Lindblad equation, and there is a large literature on stochastic methods that can be used to reduce the computational cost \cite{Dalibard:1992zz,Molmer:93,Plenio:1997ep,carmichael1999statistical,weissbook,Daley:2014fha}.   
One of the more popular approaches is the Monte-Carlo wave-function or Quantum Trajectories method, originally developed for applications in quantum optics \cite{Dalibard:1992zz}. 
In \cite{Brambilla:2020qwo}, this method was applied to quarkonium phenomenology and resulted in a substantial reduction in the computational cost.  
In addition to the reduction in computational cost, another advantage of the quantum trajectories algorithm is that this method does not require truncation in angular momentum.  
As we will discuss, the most computationally costly operation that appears in the application of the quantum trajectories algorithm to the study of quarkonium suppression is the real-time solution of the Schr\"{o}dinger equation with a non-Hermitian Hamiltonian. Luckily, there is existing literature concerning optimization of this aspect of the calculation which can be directly applied to this problem \cite{Boyd:2019arx,Islam:2020gdv,Islam:2020bnp}.

In \cite{Brambilla:2020qwo}, in order to maximize code efficiency, the computations were performed using graphical processing units (GPUs). 
In this paper, we introduce a CPU-based code in order to allow the code to be run more easily by practicing scientists. 
Associated with this paper, the CPU-based code is being made publicly available so that it can be used by others to study in-medium quarkonium suppression and, with some adjustments, similar problems. 
As detailed below, we achieve high efficiency by combining the quantum trajectories algorithm with a split-step pseudo-spectral method to compute the evolution of the wave-function subject to a non-Hermitian Hamiltonian.
We also introduce several optimizations of the code which make the CPU code comparable in speed to the GPU code.
We denote the resulting code \qtraj\!.

The paper is organized as follows. In Section \ref{sec:downloadandcompile}, we describe how to download, compile, and run \qtraj\!. In Section \ref{sec:params}, we detail the runtime parameters of the code. In Section \ref{sec:algorithm}, we provide a brief theoretical background and describe the algorithm as applied to the Lindblad equation studied in \cite{Brambilla:2020qwo}. Key parts of the code are discussed in Section \ref{sec:key}. In Section \ref{sec:notebooks}, we provide instructions for how to analyze \qtraj output. In Section \ref{sec:tests}, we present tests of the code and perform scaling analyses to determine the dependence on the parameters such as lattice size, etc. In Section \ref{sec:benchmarks}, we benchmark the code and compare to matrix-based methods for solving the Lindblad equation. Finally, in Section \ref{sec:conclusions}, we present our conclusions.  In the appendices we collect a list of required packages, details concerning the input/ouput file formats, and the potentials which are implemented.

\section{Download and compilation of the code}
\label{sec:downloadandcompile}

\subsection{How to download the code}
\label{subsec:download}

To download the most up-to-date version of \qtraj\!, the user can either pull from the public GIT repository
\begin{verbatim}
    $ git clone https://bitbucket.org/michael_strickland/qtraj-fftw.git
\end{verbatim}
or download tagged versions using the URL contained in~\cite{qtraj-download}.

\subsection{Structure of the \qtraj package}
\label{sec:structure}

After download, the \qtraj package main folder has the following top-level directory structure
\listingbox{%
\dirtree{%
.1 qtraj-fftw.
.2 build.
.2 docs.
.2 input.
.2 license.
.2 mathematica.
.2 scripts.
.2 test.
}
}

\subsection{Compilation of \qtraj and unit tests}
\label{subsec:finalcompilation}

Once the requisite packages listed in \ref{sec:requirements} have been installed, \qtraj can be compiled by simply typing
\begin{verbatim}
    $ make
\end{verbatim}
from the main directory.  The user can compile with all cores on their system by adding `{\tt -j}' or specify a given number of cores by adding `{\tt -j \#}'.  Upon successful compilation, an executable called {\tt qtraj} will be generated in the main folder.  The Makefile will automatically detect if the user has MKL installed.  If MKL is installed, the MKL version of the internal eigensolver will be used, otherwise, the Armadillo version will be used.  To reset the build, the user can execute 
{\tt make clean} from the main directory. Note that it is possible to compile the code without an internal eigensolver by executing {\tt make mode=lightweight} from the main directory.  In this case, runtime numerical determination of the eigenstates is not possible; however, the user can read previously computed eigenstates from disk (see Sec.~\ref{sec:eigensolvers} for more details).

After compiling the main executable, the user can run unit tests on the code by executing
\begin{verbatim}
    $ make tests
\end{verbatim}
from the main directory.  Note that this functionality requires the GoogleTest framework to be installed.  See \ref{sec:requirements} for further information.

To run \qtraj using the parameters from {\tt input/params.txt}, the user executes
\begin{verbatim}
    $ ./qtraj
\end{verbatim}
in the directory where the code was compiled.  
All output from the code will appear either in a directory called {\tt output} or {\tt output-<random-seed>}, depending on the output parameter settings.  

\section{Code parameters}
\label{sec:params}

All runtime parameters for the code can be set by editing the file {\tt input/params.txt}.  We note that \qtraj {\tt 1.0} does not implement consistency checks on the user supplied values of the runtime parameters.  Parameter values set from the file can be overridden by appending a dash followed by the name of the parameter and the value to use, e.g.
\listingbox{%
\tt
\$ ./qtraj -nTrajectories 16
}
sets the value of {\tt nTrajectories} to 16, overriding the value indicated in the file {\tt input/params.txt}.  In Table \ref{tab:params}, we provide documentation for each runtime parameter.\footnote{Some of the parameters describing the two KSU potentials are hard-coded; their description can be found in \ref{sec:isoksupotential} and \ref{sec:anisoksupotential}.}\\

\begin{table}
\begin{center}
{\small 
\begin{tabular}{|p{0.1\linewidth}|p{0.18\linewidth}|p{0.15\linewidth}|p{0.42\linewidth}|}
\hline
\cellcolor{gray!10} \centering {\bf Group} & \cellcolor{gray!10} \centering {\bf Parameter} & \cellcolor{gray!10} \centering {\bf Value/Type} & \cellcolor{gray!10} {\bf Description} \\
\cline{1-4}
\centering Potential&\centering potential&\centering 0&Munich potential (\ref{sec:munichpotential})\\
\cline{3-4}
\centering parameter&&\centering 1&Isotropic KSU potential (\ref{sec:isoksupotential}) \\ 
\cline{3-4}
&&\centering 2&Anisotropic KSU potential (\ref{sec:anisoksupotential})\\ 
\cline{1-4}
\centering Jump&\centering doJumps& \centering 1&Perform quantum jumps; only for potential=0\\
\cline{3-4}
\centering  parameters&&\centering 0&Do not perform quantum jumps\\
\cline{2-4}
&\centering maxJumps& \centering integer $\geq0$ &Maximum number of quantum jumps\\
\cline{1-4}
\centering Physics&\centering m&\centering real $\geq0$ &Reduced mass in units of GeV\\ 
\cline{2-4}
\centering parameters&\centering alpha&\centering real $\geq0$&Coulomb coupling; $\alpha = C_F \alpha_s $\\
\cline{2-4}
&\centering kappa&\centering -1&Central fit (\ref{sec:munichpotential}) \\
\cline{3-4}
&&\centering -2&Lower fit (\ref{sec:munichpotential}) \\
\cline{3-4}
&&\centering -3&Upper fit (\ref{sec:munichpotential})\\
\cline{3-4}
&&\centering real $\geq0$& Coefficient $\hat\kappa$ in Munich potential (\ref{sec:munichpotential})\\
\cline{2-4}
&\centering gam&\centering real $\leq$ 0 & Coefficient $\hat\gamma$ in Munich potential (\ref{sec:munichpotential})\\
\cline{1-4}
\centering Temperature&\centering temperatureEvolution&\centering 0& Ideal Bjorken evolution; $T(\tau) = T_0 (\tau_{\rm med}/\tau)^{1/3}$\\
\cline{3-4}
\centering  parameters&&\centering 1& Read from temperature evolution file (\ref{sec:formats1})\\
\cline{3-4}
&&\centering 2& Read from ``trajectory file'' (\ref{sec:formats2})\\
\cline{2-4}
&\centering temperatureFile&\centering string&Specifies temperature file path \\
\cline{2-4}
&\centering T0&\centering real $>$ 0  &Initial temperature in units of GeV (Bjorken)\\
\cline{2-4}
&\centering Tf&\centering 0 $<$ T0 $<$ Tf  & Final temperature in units of GeV\\
\cline{2-4}
&\centering tmed&\centering real $\geq0$& Time to turn on hydro background in units of GeV$^{-1}$\\
\cline{2-4}
&\centering t0&\centering real $\geq0$& Time to begin vacuum evolution in units of GeV$^{-1}$\\
\cline{1-4}
\centering Initial&\centering initType&\centering 0&Singlet Coulomb eigenstates\\
\cline{3-4}
\centering condition&&\centering 1& Gaussian delta function\\
\cline{3-4}
\centering parameters&&\centering 100& Computed eigenstates (Section \ref{sec:eigensolvers}) \\
\cline{3-4}
&&\centering 200 & Loads eigenstates from file (Section \ref{sec:eigensolvers}) \\
\cline{2-4}
&\centering ProjType&\centering 0&Uses coulomb eigenstates\\
\cline{3-4}
&&\centering 1&Uses computed eigenstates\\
\cline{3-4}
&&\centering 2&Uses disk-based eigenstates (Section \ref{sec:eigensolvers})\\
\cline{2-4}
&\centering initWidth&\centering real $\geq0$& Initial width for Gaussian IC (initType=1)\\
\cline{2-4}
&\centering initN&\centering 1&Principal quantum number $n$; only for initType=0\\
\cline{2-4}
&\centering initL&\centering 0 &Angular momentum quantum number $l$\\
\cline{2-4}
&\centering initC&\centering 0&Singlet color configuration\\
\cline{3-4}
&&\centering 1&Octet color configuration; only for potential=0\\
\cline{2-4}
&\centering basisFunctionsFile&\centering string& Specifies basis functions file path (Section \ref{sec:eigensolvers}) \\
\cline{1-4}
\centering Simulation&\centering nTrajectories&\centering integer $\geq$ 0 & Number of quantum trajectories to simulate\\
\cline{2-4}
\centering parameters&\centering randomseed&\centering 0& Uses high-resolution system timer as a seed\\
\cline{3-4}
&&\centering integer $>$ 0& Uses specified number as a seed\\
\cline{3-4}
&&\centering -1& Uses deterministic random numbers (for testing)\\
\cline{2-4}
&\centering rMax&\centering 0 $<$ real $\leq$ 1 & Sets maximum initial random number (Section \ref{ssec:algo}) \\
\cline{2-4}
&\centering maxJumps&\centering integer $\ge$ 0 & Sets maximum number of jumps (Section \ref{ssec:algo}) \\
\cline{1-4}
\centering Grid&\centering num&\centering integer $\geq$ 2 & Number of lattice sites; $2^n$ for best performance \\
\cline{2-4}
\centering parameters&\centering L&\centering real $>$ 0&Size of simulation box in units of GeV$^{-1}$\\
\cline{2-4}
&\centering dt&\centering real $> 0$ & Time step in units of GeV$^{-1}$\\
\cline{2-4}
&\centering maxSteps&\centering integer $\geq0$& Maximum number of time steps\\
\cline{2-4}
&\centering derivType&\centering 0&Full derivative  (Section \ref{sec:derivativetypes}) \\
\cline{3-4}
&&\centering 1&Second-order differences  (Section \ref{sec:derivativetypes}) \\
\cline{1-4}
\centering Output&\centering snapFreq&\centering integer $\geq$ 0& Frequency for the summary output\\
\cline{2-4}
\centering parameters&\centering snapPts&\centering integer $\leq$ num& For best performance num/snapPts=$2^n$\\
\cline{2-4}
&\centering dirnameWithSeed&\centering 0& Output directory ``output''\\
\cline{3-4}
&&\centering 1&Output directory ``output-$<$seed$>$''\\
\cline{2-4}
&\centering saveWavefunctions&\centering 0&Turns off saving of wavefunctions \\
\cline{3-4}
&&\centering 1& Turns on saving of wavefunctions \\
\cline{2-4}
&\centering outputSummaryFile&\centering 0& Turns off output of the summary file summary.tsv\\
\cline{3-4}
&&\centering 1& Turns on output of the summary file summary.tsv\\
\hline
\end{tabular}
}
\end{center}
\vspace{-3mm}
\caption{Table of \qtraj runtime parameters.  These can be adjusted within the file {\tt input/params.txt} or on the command line (see Sec.~\ref{sec:params}).
}
\label{tab:params}
\end{table}

\subsection{Using the built-in eigensolvers}
\label{sec:eigensolvers}

\qtraj includes the ability to determine the vacuum eigenstates numerically based on the potential set using the {\tt potential} parameter.  For this purpose, it makes use of either the Intel MKL library \cite{IntelOneAPI} or the Armadillo library \cite{Armadillo} (see \ref{sec:requirements} for details).  To use this feature, the user must set {\tt initType = 100}.  Upon code execution, \qtraj first determines the eigenstates and outputs the masses of the 1S, 2S, 1P, 3S, 2P, and 1D states.\footnote{There is an internal variable {\tt nBasis} that can be adjusted to output more basis states.
The set of principal and angular momentum quantum numbers of the basis states have to be adjusted by the user before recompiling the code.} 
These eigenstates are saved to a file in the output folder called {\tt basisfunctions.tsv}. 
Once generated, this file can be used to initialize the eigenstate basis.  To do this, the user should copy the {\tt basisfunctions.tsv} file into the {\tt input} folder,  adjust the parameter {\tt basisFunctionsFile} to point to this file, and finally change the parameter {\tt initType} to 200.  This is useful for running \qtraj on machines that do not have either the MKL or Armadillo packages installed.

\subsection{Derivative types}
\label{sec:derivativetypes}

During the $H_{\rm eff}$ evolution of the wave-function, the \qtraj code can make use of two different types of kinetic energy operators.  For the case {\tt derivativeType = 0}, the momentum-space representation of the kinetic energy operator $T_\text{kin} = p^2/m$ is used, where $m$ is the mass of the heavy quark; $m/2$ is the reduced mass in the heavy quarkonium system.  This is the default setting in \qtraj\!.  If the user specifies {\tt derivativeType = 1}, the momentum-space representation of the kinetic energy operator is taken to be of the form $T_\text{discrete} = 2 [1-\cos(p\Delta r)]/(m\Delta r^2)$, which corresponds to using centered differences to compute the second derivative operator.  This option is less accurate than using the full kinetic energy operator.  It exists to facilitate comparison to codes using second order differences.  

\section{Theoretical background and description of the algorithm}
\label{sec:algorithm}

\subsection{Lindblad equation}
\label{sec:Lindblad}

For a wide range of quantum-mechanical applications, we are faced with a 
Markovian time evolution of a density matrix $\rho$ of a subsystem 
interacting with an environment that preserves the trace of the subsystem's 
density matrix. 
Such a system can be described by the GKSL or Lindblad equation
\cite{Gorini:1975nb,Lindblad:1975ef} whose general form is given by 
\begin{equation}  \frac{d\rho}{dt}=-i[H,\rho]+\sum_n 
\left(  C_n\,\rho\,C_n^\dagger-\frac{1}{2}\{C_n^\dagger C_n,\rho\} \right).
\label{eq:Lindblad}
\end{equation}
$H$ is a Hermitian Hamilton operator that describes deterministic and unitary time 
evolution of the subsystem in the presence of the environment, 
whereas the collapse operators $C_n$ describe processes involving 
explicit exchanges of quanta between the subsystem and the environment. 
On the level of the subsystem, these lead to non-unitary time evolution. 
Defining for those non-unitary interactions partial and total widths
\begin{equation}
\Gamma_n=C_n^\dagger C_n\,,\quad 
\Gamma = \sum\limits_n \Gamma_n\,,
\label{eq:width}
\end{equation}
we can reformulate the Lindblad equation in terms of a non-Hermitian effective 
Hamiltonian 
\begin{equation}
H_\text{eff}=H-\frac{i}{2}\Gamma\,,
\label{eq:Heff}
\end{equation}
that gives rise to non-unitary, but deterministic, time evolution and transitions between  
different quantum states that are driven by the term $C_n\,\rho\,C_n^\dagger$.
Rewriting the Lindblad equation as 
\begin{equation}
\frac{d\rho}{dt}=-iH_\text{eff}\rho+i\rho H^\dagger_\text{eff}+\sum_n C_n\,\rho \,C_n^\dagger\,,
\end{equation}
we are in a position to discretize the time evolution in a form that is 
in principle amenable to numerical integration, 
\begin{equation}
\rho(t+dt) = \rho(t)-iH_\text{eff}\rho(t)dt+i\rho(t)H_\text{eff}^\dagger dt+\sum_n C_n\,\rho(t) \,C_n^\dagger dt\,,
\label{eq:discLE}
\end{equation}
where each discrete step of numerical integration has a diagonal and 
an off-diagonal component. 
Due to the linearity of Eq. \eqref{eq:Lindblad}, one may have any pure state 
as initial condition and average the late time results over different appropriately 
weighted initial states. 

\subsubsection{First principles description of quarkonium using effective field theory}
\label{sec:1stprinciple}

A heavy quark-antiquark pair coupled to a hot, nuclear medium can be described using the formalism of Open Quantum Systems (OQS)
applied to the effective field theory (EFT) potential Non-Relativistic QCD (pNRQCD); we denote this combined approach OQS+pNRQCD.
pNRQCD is obtained from QCD by first integrating out the scale of the heavy-quark mass $m$, leading 
to Non-Relativistic QCD (NRQCD) \cite{Caswell:1985ui,Bodwin:1994jh,Pineda:1997bj}, and then successively performing a multipole expansion in the 
relative distance $r$ between the heavy quark and antiquark \cite{Pineda:1997bj, Brambilla:1999xf}. The latter 
step integrates out the soft scale $1/r \sim mv$ and the resulting EFT describes heavy quark-antiquark pairs in 
color singlet or octet configurations as well as light quarks and gluons at the ultrasoft scale $\als/r \sim mv^2$. 
The Wilson coefficients of pNRQCD are functions of $r$, and play the role of potentials, although higher order contributions have non-potential form as well.
The evolution equation for the heavy quarkonium reduced density matrix can be obtained within pNRQCD and cast into Lindblad form as long 
as the medium's scales are much smaller than the inverse Bohr radius and much larger than the ground state binding 
energy \cite{Brambilla:2016wgg,Brambilla:2017zei}.  
For other hierarchies, a Lindblad form cannot be obtained. 
In anticipation of details that we review later, we mention that the problem 
has radial symmetry if the medium is isotropic. 
Hence, the internal quanta that are exchanged between the subsystem and the 
environment relate to the color configuration (singlet or octet) of the heavy 
quark-antiquark pair and its angular momentum; the deterministic evolution 
with $H_\text{eff}$ leaves the color state as well as the angular momentum 
invariant, while the collapse operators change color and angular momentum in 
a well defined manner.
For a brief overview of the structure, see Section \ref{ssec:OQSpNRQCD}; for 
hard-coded numerical values of the parameters, see \ref{sec:munichpotential}.

\subsubsection{Phenomenological descriptions of quarkonium}
\label{sec:pheno}

Alternatively, one may use a hard-coded, phenomenologically motivated interaction 
of the heavy quark-antiquark pair.
A KMS potential ~\cite{Karsch:1987pv, Dumitru:2009ni, Strickland:2011aa} 
with an imaginary part as dictated by the next-to-leading order result in the 
electric screening regime ~\cite{Laine:2006ns} is provided in its isotropic 
(see \ref{sec:isoksupotential}) or anisotropic (see \ref{sec:anisoksupotential}) variants. 
These phenomenological potentials consider only the color singlet and deterministic 
evolution with $H_\text{eff}$; there are no collapse operators implemented in the phenomenological models included in this version. 
We note that it is straightforward to extend the code to include new phenomenological potentials. In addition, it is possible for users to include collapse terms for phenomenologically motivated potentials.

\subsection{The Monte-Carlo Wave-Function method}
\label{ssec:MCWF}

Direct evaluation of Eq. \eqref{eq:discLE} is very demanding and necessarily approximate. 
Besides the obvious need for a sufficiently fine discretization of time, a 
discretization of the underlying space on a grid with a finite spacing and 
a finite box size (or decompostition in terms of a finite number of basis functions) 
is required. 
If the interaction is isotropic and central, it is sufficient to discretize the 
radial direction and treat angular momentum as an internal quantum number. 
Any numerical implementation of $\rho(t)$ necessitates that the Hilbert space be 
truncated to a finite set of values of each internal quantum number among which  
the system can fluctuate through successive applications of the collapse operators. 
In \cite{Brambilla:2016wgg,Brambilla:2017zei},  Eq. (\ref{eq:discLE})
was  solved numerically by considering only $S$- and $P$-wave states. 
Because the size of the density matrix grows as $N^2$, where $N$ is the product of 
the number of spatial grid points and of the number of internal quantum states, 
extending the cutoffs quickly makes the numerical integration prohibitively numerically expensive. 

As usual, the only known escape from this curse of dimensionality is stochastic
integration. 
In the literature, various stochastic techniques have been developed to tackle 
this issue, and they go by the name of \textit{master equation unravelling} (see 
\cite{Breuer:2002pc} and references therein). 
One such unravelling, the Quantum State Diffusion \cite{Gisin:1992xc}, has been 
already applied to the study of quarkonium, see for example the 
recent papers \cite{Miura:2019ssi,Sharma:2019xum}. 

We use a variant of the Monte-Carlo Wave-Function (MCWF) method \cite{Dalibard:1992zz} that is particularly advantageous for compact in-medium heavy quark-antiquark pairs 
and makes efficient use of the symmetries of their interactions. 
In particular, $H_\text{eff}$ is diagonal in the space of internal quantum numbers. 
We write the density matrix as a linear combination of projection operators 
$\rho(t) = \sum_{\psi} p_\psi(t) P_\psi$, $P_\psi=\ket{\psi} \bra{\psi}$, 
where $\psi$ is a multi-index characterizing the quantum state; in the case of quarkonium, 
the angular momentum quantum number $l=0,1,2,\ldots$, the internal quantum number 
(color $c$, which can be singlet $s$ or octet $o$ which represents the eight equivalent color octet states), and the radial dependence must be specified.
Assuming azimuthal symmetry of the isotropic medium and the interaction, we can 
integrate over the azimuthal angle and omit the magnetic quantum number $m$. 
$p_\psi(t) \ge 0$ are the probabilities for finding the system at time $t$ 
in the state $\ket{\psi(t)}$ and satisfy $\sum_\psi p_\psi(t) =1$. 

As mentioned above, any state of the subsystem at any time $t$ can be understood 
as a superposition of initial pure states that have been evolved to time $t$. 
For this reason, it is sufficient to consider Eq. \eqref{eq:discLE} for the evolution 
of a pure state $\ket{\psi(t)}$ with density matrix 
$\rho(t) = \ket{\psi(t)}\bra{\psi(t)}$ 
by one time step $dt$:
\begin{align}
\rho(t+dt) 
& = (1-iH_\text{eff}dt)\ket{\psi(t)}\bra{\psi(t)}
  +i\ket{\psi(t)}\bra{\psi(t)}H_\text{eff}^\dagger dt
  +\sum_n C_n\,\ket{\psi(t)}\bra{\psi(t)} \,C_n^\dagger dt\nonumber \\
& = (1-\bra{\psi(t)}\Gamma\ket{\psi(t)} dt) \, 
  \frac{(1-iH_\text{eff}dt)\ket{\psi(t)}\bra{\psi(t)}(1+iH_\text{eff}^\dagger dt)}
  {(1-\bra{\psi(t)}\Gamma\ket{\psi(t)} dt)} 
\nonumber \\& \hspace{1cm}
  + \sum_n\langle\psi(t)|\Gamma_n|\psi(t)\rangle dt \, \frac{C_n\,\ket{\psi(t)}\bra{\psi(t)} \,C_n^\dagger dt}{\langle\psi(t)|\Gamma_n|\psi(t)\rangle dt}+\mathcal{O}(dt^2)\,.
\label{eq:evol}
\end{align}
The expectation value of the total width $\Gamma$ for the time evolved state $\ket{\psi(t)}$ at time $t$, 
\begin{equation}
\Gamma(t) \equiv \bra{\psi(t)}\Gamma\ket{\psi(t)} \ge 0 \, ,    
\end{equation}
can be used to define probabilities for two different modes of evolution in a single time step\footnote{
$dt$ can always be chosen small enough such that $\Gamma(t) dt \le 1$}, either by deterministic, 
non-unitary evolution, or by a quantum jump. 
The stochastic pattern of both types of intermittent evolution characterizes each quantum trajectory $\ket{\psi(t)}$.

How many such trajectories $\ket{\psi(t)}$ are required to achieve sufficient statistical coverage 
of the ensemble strongly depends on the problem in question; see Section \ref{sec:benchmarks} for a discussion 
of the required number of trajectories.
Although the number of necessary trajectories may be large for any generic observable, there are usually 
many observables of interest for which importance sampling can be applied in a rather straightforward 
fashion to obtain accurate predictions with a much smaller number of trajectories. 
The MCWF method may offer computational advantages compared to the direct evolution of the density matrix. 
First and foremost, the trajectories are completely independent of each other and their computation is 
embarrassingly parallel. 
Second, the problem size for computing any individual trajectory scales only linearly and only with the 
number of spatial grid points, but not with the internal quantum numbers. 
Although the number of angular momentum or internal quantum states influences the necessary number of trajectories, importance 
sampling techniques or physically sensible restrictions on the simulation parameters could have a strong limiting effect. 
Lastly, the MCWF method does not require an artificial truncation of the angular momentum or internal quantum numbers. 

\subsubsection{The waiting time approach}
\label{ssec:WTA}

The main drawback of the naive implementation of the MCWF evolution given in Eq. \eqref{eq:evol} is that jumps take a finite amount of time $dt$ equal to one step in the time discretization of the deterministic evolution with $H_\text{eff}$. 
As such, the time between jumps is systematically underestimated. 
We use instead the waiting time approach (see \cite{Daley:2014fha} and references therein) in the numerical implementation of the MCWF method, since it treats the quantum jumps as instantaneous. 
We always consider a normalized, pure initial state $\ket{\psi(t_i)}$ in the 
waiting time approach. 
The probability $p_\text{Jump}(t_{i+1},t_{i+1}+\delta t)$ that the next consecutive jump 
occurs in the interval $[t_{i+1},t_{i+1}+\delta t]$ is given by the difference $p_\text{NoJump}(t_{i+1})-p_\text{NoJump}(t_{i+1}+\delta t)$ where $p_\text{NoJump}(t)$ is the probability of no jump occurring until time $t$.
This can be linearized in terms of the jump rate 
$R_\text{Jump}(t_{i+1})$ as 
\begin{align}
\int_{t_{i+1}}^{t_{i+1}+\delta t} \hskip-2em dt^\prime  R_\text{Jump}(t^\prime)
&=R_\text{Jump}(t_{i+1})\, \delta t + \mathcal{O}(\delta t^2)
= p_\text{Jump}(t_{i+1},t_{i+1}+\delta t) 
=  p_\text{NoJump}(t_{i+1}) -p_\text{NoJump}(t_{i+1}+\delta t) 
\nonumber\\
&= -\frac{d p_\text{NoJump}(t_{i+1})}{d t_{i+1}}\, \delta t + \mathcal{O}(\delta t^2)\,.
\end{align}
The probability that no jump has occurred and only deterministic, non-unitary evolution with $H_\text{eff}$ 
has taken place from time $t_{i}$ to $t_{i+1}$ is given by the norm of the state evolved just with 
$H_\text{eff}$ to time $t_{i+1}$ 
\begin{equation}
p_\text{NoJump}(t_{i+1}) 
= \abs{\ket{\psi(t_{i+1})}}^2
= \bra{\psi(t_i)} e^{+i\int_{t_{i}}^{t_{i+1}}dt^\prime H_\text{eff}^\dagger} 
e^{-i\int_{t_{i}}^{t_{i+1}} dt^\prime H_\text{eff}}\ket{\psi(t_i)}, 
\quad 
\ket{\psi(t_{i+1}) } \equiv e^{-i\int_{t_i}^{t_{i+1}}\,dt'H_\text{eff}(t')} \ket{ \psi(t_i)}. 
\label{eq:wta evol}
\end{equation}
We obtain the jump rate $R_\text{Jump}(t_{i+1})$ at time $t_{i+1}$ as the derivative of the norm. 
This can be related to the expectation value of the width $\Gamma$ for the state at time $t_{i+1}$ by 
taking the derivative of Eq. \eqref{eq:wta evol},
\begin{align}
R_\text{Jump}(t_{i+1}) 
= - \frac{d p_\text{NoJump}(t_{i+1}) }{d t_{i+1}}
= -\frac{d \abs{ \ket{ \psi(t_{i+1})} }^2 }{d t_{i+1}}
= \bra{\psi(t_{i+1})} \Gamma \ket{ \psi(t_{i+1})}
= \Gamma(t_{i+1}).
\label{eq:wta}
\end{align}
The instantaneity of the jumps is the main advantage of the waiting time approach, i.e. the time 
interval size $\delta t$ that is connected to the jumps can be chosen independently from the time interval $dt$ that 
is related to the evolution with $H_\text{eff}$ and thus determines the numerical error due to the finite time discretization.

A simple way to obtain the distribution in Eq.~\eqref{eq:wta} is to generate one uniformly distributed 
random number $r_i \in [0,1]$ for each $t_i$ and evolve with $H_\text{eff}$ until 
$ \abs{ \psi(t_{i+1}) }^2 \le r_i$. 
At this time, a jump occurs, and additionally, one or more random numbers $r_i^\prime \in [0,1]$ may be required to determine 
which collapse operator $C_{n}$ with probability
\begin{equation}
p_{n} = \frac{\bra{ \psi(t_{i+1})} C_{n}^{\dagger}C_{n}  \ket{ \psi(t_{i+1})}}{\sum_{n} \bra{ \psi(t_{i+1})} C_{n}^{\dagger}C_{n}  \ket{ \psi(t_{i+1})}},
\end{equation}
is applied in the jump.
The jumped state at time $t_{i+1}$ is normalized again and becomes the new initial state,
\begin{equation}
\ket{ \psi(t_{i+1})} 
= \frac{ C_n\ket{ \psi(t_{i+1})}}{ \big|\big| C_n\ket{ \psi(t_{i+1})}\big|\big|}.
\end{equation}
A new random number $r_{i+1}\in [0,1]$ is drawn and the cycle restarts with $\ket{\psi(t_{i+1}) }$ 
as the initial state deterministically time evolved with $H_\text{eff}$ until a jump is triggered or the simulation end time is reached. 

\subsubsection{Algorithm for deterministic evolution between quantum jumps}
\label{sec:splitstep}

For the deterministic evolution of the wave-function between jumps, we use a 
split-step pseudo-spectral or Strang splitting method \cite{Fornberg:1978,TAHA1984203,Boyd:2019arx}. 
Due to the radial symmetry of the underlying potentials, one may compute 
the deterministic evolution of a state with any given angular momentum $l$ and 
internal (color $c$) quantum number using 
\begin{equation}
u(r,t + dt) = \exp(- i H_{\text{eff}} dt) u(r,t) \, ,
\label{eq:uUpdate}
\end{equation}
where $u(r,t) = rR(r,t)$ with $R(r,t)$ being the radial part of the wave-function.  
The effective Hamiltonian $H_\text{eff}$ as defined in Eq.~\eqref{eq:Heff} contains the angular momentum term and a 
potential $V(r)$ which depends on the internal quantum number(s).  
For the normalization of the wave-function, we take
\begin{equation}
\frac{\texttt{L}}{\texttt{NUM}}\sum_{r=1}^{\texttt{NUM}} u^*(r,t) u(r,t) = 1 \, .
\label{eq:normcond}
\end{equation}
We enforce the boundary condition $u(r=0,t)=0$ through a real-valued 
Fourier sine series in a domain $r \in (0,{\texttt{L}}]$ to describe both the real 
and imaginary parts of the wave-function. 
To perform the update specified in Eq.~\eqref{eq:uUpdate}, we split the Hamilton operator 
into its kinetic and effective potential parts $H_\text{eff} = T_\text{kin} +  V_\text{eff}$; 
the entire dependence on angular momentum $l$ and internal (color $c$) quantum numbers and the non-hermiticity are in the effective potential term $V_\text{eff}$.
We use the Baker-Campbell-Hausdorff theorem to approximate
\begin{equation}
\exp\left[- i H_\text{eff} dt\right] \simeq \exp\left[- i V_\text{eff} \frac{dt}{2} \right] \exp\left[- i T_\text{kin} dt\right]  \exp\left[- i V_\text{eff} \frac{dt}{2} \right] + \mathcal{O}(dt^2) \, .
\end{equation}
The resulting sequence for a single time step of deterministic evolution is: 
\begin{enumerate}
	\item Update in configuration space using a half-step: $ \psi_1 = \exp\left[- i V_\text{eff} \frac{dt}{2} \right] \psi_0$.
	\item Separately Fourier sine transform real and imaginary parts: $\tilde\psi_1 = \mathbb{F}_s[\Re \psi_1] + i \mathbb{F}_s[\Im \psi_1] $.
	\item Update in momentum space using: $\tilde\psi_2 =  \exp\left[- i T_\text{kin} dt\right] \tilde\psi_1$.
	\item Separately inverse Fourier sine transform real and imaginary parts: $\psi_2 = \mathbb{F}_s^{-1}[\Re \tilde\psi_2] + i \mathbb{F}_s^{-1}[\Im \tilde\psi_2] $.
	\item Update in configuration space using a half-step: $ \psi_3 = \exp\left[- i V_\text{eff} \frac{dt}{2} \right] \psi_2$.
\end{enumerate}
For real valued potentials, the Strang splitting leads to manifestly unitary evolution.
In order to evaluate operators in momentum space, we use the Fourier sine 
transform ($\mathbb{F}_s$) and inverse Fourier sine transform 
($\mathbb{F}_s^{-1}$) provided by the FFTW3 package \cite{FFTW}.
The default implementation of the momentum-space operator is $T_\text{kin} = p^2/m$; however, 
there is an alternative implementation of the second derivative via centered discrete 
differences, $T_\text{discrete} = 2[1-\cos(p \, dr)]/(m \, dr^2)$ (see Sec. \ref{sec:derivativetypes}).  

We note that a major benefit of the DST algorithm is that when using the full derivative operator $T_\text{kin}$ the derivatives are effectively computed using all 
points on the lattice rather than a fixed subset of points. 
As a result, the evolution obtained using the DST algorithm is more accurate than that obtained using, for example, a Crank-Nicolson (CN) scheme with a three-point second-derivative \cite{Boyd:2019arx}.  
Using the same sized derivative stencil as the default DST algorithm, the CN scheme scales as ${\cal O}(N^2)$ where $N$ is the number of lattice points in the wave-function, whereas the split-step DST evolution scales as ${\cal O}(N \log N)$~\cite{Boyd:2019arx}.

\subsection{In-medium quarkonium evolution in OQS+pNRQCD}
\label{ssec:OQSpNRQCD}

In the following we review the ingredients of the master equation obtained from OQS+pNRQCD 
\cite{Brambilla:2016wgg,Brambilla:2017zei} at order $\mathcal{O}(r^2)$ and leading order in $E/T$, i.e. for compact, Coulombic quarkonia in the regime $1/a_0 \gg T,m_D\gg E$, for which the master 
equation has Lindblad form.

\subsubsection{Hamiltonian in OQS+pNRQCD }

In this regime, the Hamiltonian at order $\mathcal{O}(r^2)$ is given as 
\begin{equation}
H = \left(\begin{array}{c c}
h_s & 0\\
0 & h_o
\end{array}\right)
+ \frac{r^2}{2}\,\gamma\, 
\left(\begin{array}{c c}
1 & 0\\
0 & \frac{N_c^2-2}{2(N_c^2-1)}
\end{array}\right)\,,
\label{eq:hamiltonian}
\end{equation}
where $h_{s,o}$ is the singlet, octet in-vacuum Hamiltonian in pNRQCD \cite{Pineda:1997bj,Brambilla:1999xf}:
\begin{equation}
h_{s,o} = \frac{{\bf p}^2}{m} +V_{s,o}(r)\,.
\label{eq:hsoCoul}
\end{equation}
Since we assume that the quarkonium is a Coulombic bound state, the potential $V_{s,o}(r)$ in Eq. \eqref{eq:hsoCoul} 
is the attractive, singlet Coulomb potential $V_s(r) =-C_F\, \als/r$ where $C_F=(N_c^2-1)/(2N_c)$ or the repulsive, octet Coulomb potential $V_o(r) =+\als/(2N_c r)$. 
In writing Eq. \eqref{eq:hamiltonian}, we have assumed a component structure of the quarkonium state 
as $ \ket{ \psi} = ( \ket{S} , \ket{O} )^T$. 

The coefficient $\gamma$ of the order $\mathcal{O}(r^2)$ term in Eq. \eqref{eq:hamiltonian} is the dispersive 
counterpart of the temperature dependent heavy-quark momentum diffusion coefficient $\kappa$. 
It enters by matching pNRQCD with QCD.

\subsubsection{Collapse operators, partial widths, and color state transitions in OQS+pNRQCD}

There are six collapse operators: $C^0_i$ and $C^1_i$ (the spatial index $i$ 
corresponds to the spatial directions and assumes values 1,2,3).
The $C^0_i$ induce transitions between singlet and octet states, while the $C^1_i$ induce 
transitions among octet states.
They read
\begin{equation}
C^0_i=\sqrt{\frac{\kappa}{N_c^2-1}}\,r^i\left(\begin{array}{c c}
0 & 1\\
\sqrt{N_c^2-1} & 0
\end{array}\right)\,,\quad
C^1_i=\sqrt{\frac{(N_c^2-4)\kappa}{2(N_c^2-1)}}\,r^i\left(\begin{array}{c c}
0 & 0\\
0 & 1
\end{array}\right)\,.
\label{eq:collapse}
\end{equation}
The coefficient $\kappa$ is the heavy-quark momentum diffusion 
coefficient ~\cite{CasalderreySolana:2006rq,CaronHuot:2007gq};
like its dispersive counterpart $\gamma$, it enters by matching pNRQCD with QCD.

Two partial decay widths (cf. Eq. \eqref{eq:width}) are defined through the collapse operators
\begin{equation}
\Gamma^0=\sum_i \kappa r^i\left(\begin{array}{c c}
1 & 0\\
0 &\frac{1}{N_c^2-1}
\end{array}\right)r^i\,,\quad
\Gamma^1=\sum_i \frac{\kappa(N_c^2-4)}{2(N_c^2-1)}r^i\left(\begin{array}{c c}
0 & 0\\
0 & 1
\end{array}\right)r^i\,.
\end{equation}
Their sum gives the total decay width,
\begin{equation}
\Gamma=\sum_i \kappa r^i\left(\begin{array}{c c}
1 & 0\\
0 & \frac{N_c^2-2}{2(N_c^2-1)}\end{array}\right)r^i\,.
\end{equation}

The transition probabilities are more easily obtained in the color than in the angular momentum sector.
Singlet states can only jump to octet states: $p_{s \to o}= 1$. 
Octet states jump to octet states with probability $p_{o \to o}=(N_c^2-4)/(N_c^2-2)$\footnote{
This ratio is obtained by comparing the expectation value of the partial width 
$\Gamma^1$ to the that of the total width for any octet state.} and to singlet states with probability $p_{o\to s}=2/(N_c^2-2)$.

\subsubsection{Structure of the density matrix}

Let us examine the general structure of the density matrix $\rho$ for any operator $H_\text{eff}$ 
that is diagonal (i.e. depends only on $r=\sqrt{\bm{r}^2}$) in the angular momentum quantum 
number $l$  and all internal quantum numbers $c$ (i.e. color in OQS+pNRQCD). 
The radial and angular parts of the wave-functions can be separated multiplicatively; the 
latter is given in terms of spherical harmonics. 
Any density matrix can be expanded on the sphere in terms of spherical harmonics as 
\begin{equation}
\rho_{lm;l^\prime m^\prime } \propto \int\,d\Omega(\hat{r})\,d\Omega(\hat{r}^\prime ) \, Y_{lm}^{\phantom{\ast}}(\hat{r}) \, \rho \, Y_{l^\prime m^\prime }^\ast(\hat{r}^\prime )\,.
\end{equation}
Since there is azimuthal symmetry, the magnetic quantum number $m$ can always be summed over, 
and we can define a reduced density matrix $\rho_{l;l^\prime }$ that is independent of $m$. 
Since the effective Hamiltonian is diagonal in angular momentum and internal quantum numbers, 
the components of the density matrix are characterized as $\rho = ( \rho_{cl}(r) )$, where 
$c$ indicates the internal (color) quantum number and 
$l=0,1,2,\ldots$ indicates the angular momentum quantum number.  
We understand $\mathrm{Tr}\,[\rho_{cl}(r)]$ as the probability to find the 
subsystem in any state with angular momentum squared equal to $l(l+1)$ and internal quantum 
number $c$.
The total density matrix $\rho$ of the subsystem can be cast into the block-diagonal form 
\begin{equation}
\rho=\left(\begin{array}{ccc ccc}
\rho_{s0} & 0 & 0 & 0 & 0 & 0  \\
0 & \rho_{s1} & 0 & 0 & 0 & 0 \\
0 & 0 & \ddots & 0 & 0 & 0  \\
0 & 0 & 0  & \rho_{o0} & 0 & 0  \\
0 & 0 & 0  & 0 & \rho_{o1} & 0  \\
0 & 0 & 0  & 0 & 0 &  \ddots \\
\end{array}\right)\,,
\label{eq:rhoso}
\end{equation}
which is preserved at all times if it has been realized at any time $t_0$, i.e. in the 
initial state. 
For simplicity of the following discussions, let us assume that we have a density matrix 
of states with a well-defined angular momentum and a well-defined internal quantum number, 
i.e. only a single non-vanishing block of the density matrix. 
Since the Lindblad equation is linear in $\rho$, it is trivial to generalize these considerations. 

\subsubsection{Angular momentum transitions}

In order to understand angular momentum transitions, we need to remember that the jump term is 
$\sum_{n} C_n\rho C_n^\dagger$, with $C_n$ defined in Eq.~\eqref{eq:collapse}, where $n=(c,i)$ 
is a multi-index that ranges over $i=1,2,3$ for its spatial part and $c=0,1$ for its internal 
(color) part. 
Hence, the term that contributes in the master equation involves a sum over all spatial components, 
with two powers of the same collapse operator; this results in a radially symmetric function of 
$r^2$, and preserves the block-diagonal structure of the density matrix in Eq.~\eqref{eq:rhoso}. 
Any individual jump operator $C_n$ induces a (chromoelectric) dipole transition: namely, when applied to any state with well-defined angular momentum 
quantum number $l$, $C_n$ can produce only a state with angular momentum quantum number $l^\prime = l\pm 1$.
For either internal quantum number component of the collapse operators (either value of $c=0,1$), 
we have the probability to jump to either of the (up to two) values of $l^\prime$ given by 
\begin{equation}
p(l \to l^\prime) = 
\frac{ \sum\limits_{i} \abs {\bra{ \psi_{c^\prime l^\prime}(r)} C_{i}^c \ket{ \psi_{cl}(r)} }^2 }
{ \sum\limits_{i^\prime} \sum\limits_{l^{\prime\prime}=l \pm 1} \abs {\bra{ \psi_{c^\prime l^{\prime\prime}}(r)} C_{i^\prime}^c \ket{ \psi_{cl}(r)} }^2}
=
\frac{ \sum\limits_{i} \sum\limits_{m,m^\prime} \abs { Y_{l^\prime m^\prime}^\ast r_{i} Y_{l m}^{\phantom{\ast}} }^2 }
{ \sum\limits_{i^\prime} \sum\limits_{l^{\prime\prime}=l \pm 1} \sum\limits_{m,m^{\prime\prime}} \abs {  Y_{l^{\prime\prime} m^{\prime\prime}}^\ast r_{i^\prime} Y_{l m}^{\phantom{\ast}} }^2}
=\frac{1}{(2l+1)} \sum\limits_{i} \sum\limits_{m,m^\prime} \abs { Y_{l^\prime m^\prime}^\ast \frac{r_{i}}{r} Y_{l m}^{\phantom{\ast}} }^2
,\,
\end{equation}
where the prefactor is due to normalization of spherical harmonics. 
For each $l$, the probabilities to jump to higher or lower $l$ are 
\begin{align}
p_{l,\pm} = p(l \to l \pm 1) =\frac{1}{2l+1}\sum_{m,m'} \abs{ Y_{l\pm 1\,m^\prime}^\ast \left(\frac{r_i}{r}\right) Y_{lm}^{\phantom{\ast}} }^2\,,
\label{eq:Pl+-}
\end{align}respectively, with $p_{l,+}=1-p_{l,-}$. 
We generally have $p_{l,-}=\frac{l}{2l+1} < p_{l,+}=\frac{l+1}{2l+1}$, which is straightforward to obtain from recurrence relations. 
As jumping to higher angular momentum is always favored, 
equilibration is not reached before $l=\infty$. 
This is not surprising, as the evolution in the set of $l=0,1,2,\ldots$ must be a directed random walk 
away from the lower bound.

\subsubsection{Structure of the jump probability matrix}

Combining both observations, we can sketch the structure of the jump probability matrix as 
\begin{equation}
p_{jump}=\left(\begin{array}{cc}
0                         &    \frac{2}{N_c^2-2}     p_{l\to l^\prime} \\
p_{l\to l^\prime} & \frac{N_c^2-4}{N_c^2-2} p_{l\to l^\prime} \\
\end{array}\right)\,,
\quad
p_{l\to l^\prime} =\left(\begin{array}{ccccc}
0           & \frac{1}{3} & 0           & 0           & \hdots\\
1 & 0           & \frac{2}{5} & 0           & \hdots \\
0           & \frac{2}{3} & 0           & \frac{3}{7} & 0\\
0           & 0           & \frac{3}{5} & 0           & \ddots\\
\vdots      & \vdots      & 0           & \ddots      & \ddots \\
\end{array}\right)\,.
\label{eq:pjump}
\end{equation}
The angular momentum probability matrix $p_{l\to l^\prime}$ is infinitely large. 
We generate independent random numbers for the angular momentum and the color transitions.

Note that there are no jump matrices for the particular phenomenological potentials 
of \ref{sec:isoksupotential} and \ref{sec:anisoksupotential} since these interactions do not 
permit transitions that are off-diagonal in angular momentum or internal quantum numbers. 
However, the preceding considerations can be straightforwardly extended to arbitrary forms of 
interactions with jump matrices that lead to radially symmetric partial widths.

\subsubsection{Hydrodynamic medium evolution with time and temperature dependent potential}

The entirety of the in-medium effects in OQS+pNRQCD in the regime $1/a_0 \gg T,m_D\gg E$ at order $\mathcal{O}(r^2)$ in the multipole expansion and leading order in $E/T$ is contained in the coefficients $\kappa$ and $\gamma$. 
Changes in the hot, nuclear medium during the lifetime of the QGP induce changes in $\kappa$ and $\gamma$.
For a locally equilibrated medium such that a temperature can be defined (locally), 
the full temperature dependence of the master equation follows from these two coefficients;
on dimensional grounds, they are expected to scale as $\kappa \equiv \hat{\kappa} T^3$ and $\gamma \equiv \hat{\gamma} T^3$, 
where $\hat{\kappa}$ and $\hat{\gamma}$ are dimensionless numbers that may have further, nontrivial temperature dependence. 
Details of the time evolution of the medium and its temperature rely on a hydrodynamical description 
of the medium; for simplified benchmarks and code tests, Bjorken evolution can be used, i.e. $T(\tau) = T_0 (\tau_{\rm med}/\tau)^{1/3}$ (see Table \ref{tab:params}).
Results for $\kappa(T)$ and $\gamma(T)$ have been obtained from lattice gauge theory 
simulations in local thermal equilibrium ~\cite{Brambilla:2019tpt, Brambilla:2020siz}; 
however, some modeling of their temperature dependence constrained by the lattice results is still necessary.
Such lattice based models coupled to a hydrodynamical evolution of the medium could, therefore, describe 
how these heavy-quark momentum diffusion coefficients evolve in time.

\subsection{Summary of \qtraj execution}
\label{ssec:algo}

We collect the relevant information and enumerate a practical algorithm taking the form of a multi-loop structure.

\begin{enumerate}
    \item
    The outermost loop (scripts external to the actual \qtraj code) is over possible physics scenarios. 
    This includes varying \texttt{Grid}, \texttt{Physics}, 
    \texttt{Potential}, and \texttt{Temperature parameters}, e.g. Bjorken 
    evolution or different hydrodynamical evolutions encoded by different 
    ``temperature evolution'' or ``trajectory files'' (see \ref{sec:formats}). 
    The final time of each trajectory's evolution is determined by the time at which the temperature along the trajectory drops below $T_f$ after which no further evolution is necessary since below $T_f$ the vacuum potential is used.
    \item
    The next loop (still external to the \qtraj code) is over initial states of the subsystem. 
    Its density matrix is a linear combination of pure states $\ket{ \psi_{cl}(0) }$ 
    with well-defined angular momentum ($l$) and internal ($c$) quantum numbers
    \begin{equation}
        \rho(0) = \sum_{\psi_{cl}} p_{\psi_{cl}} \ket{ \psi_{cl}(0) } \bra{ \psi_{cl}(0) }.
    \end{equation}
    We select from among the states $\ket{ \psi_{cl}(0) }$ with the appropriate 
    \emph{initial state probability} $p_{\psi_{cl}}$.
    In the current implementation, the initial state is typically a vacuum eigenstate or a Gaussian multiplied by an appropriate 
    power of $r$ to realize the desired angular momentum in the initial state; such initial states can be 
    color singlet or octet.
    \item \label{step:no_jumps}
    The actual \qtraj simulation code begins here. 
    We first compute one quantum trajectory with in-vacuum evolution of $\ket{ \psi_{cl}(0) }$ 
    until the onset of coupling to the medium at $t_0$; the result $\ket{ \psi_{cl}(t_0) }$ 
    is stored and reused later. 
    For each quantum trajectory, we define $\ket{ \psi_{cl}(t_{i}) } \equiv \ket{ \psi_{cl}(t_0) }$ 
    for $i=0$. 
    The zeroth trajectory completes without jumps, and its final state overlaps are 
    recorded.  The corresponding final state norm equals the probability $p_\text{NoJumps}$ of no jumps occurring over the entire course of the evolution.
    The outermost loop of \qtraj is over a set of \texttt{nTrajectories} quantum trajectories for 
    the given initial condition. 
    \item \label{step:jumps}
    Each quantum trajectory is computed in a separate loop over the evolution time, starting with 
    the current state of the random number generator (as after the previous quantum trajectory);
    the code alternates between intermittent, iterative $H_\text{eff}$ evolution and 
    instantaneous jumps for each individual quantum trajectory.
    \begin{enumerate}
        \item 
        We draw a random number $r_i \in [0,1]$; 
        in the first step of the first trajectory $i=0$, we distinguish two cases: 
        \begin{enumerate}
            \item
            If $r_0 < \mathtt{rMin} \equiv p_\text{NoJumps}$, we terminate the quantum trajectory and 
            default to the known no-jump final state computed in step~\ref{step:no_jumps}.
            \item
            If $r_0 > \mathtt{rMax}$ (see Table \ref{tab:params}) we terminate the quantum trajectory 
            and report total loss for all low-lying bound states (zero overlaps).
        \end{enumerate}
        In the innermost loop, we evolve the state $\ket{ \psi_{cl} (t_i) }$ with $H_\text{eff}$ 
        to time $t_{i+1}$ while $ r_i < \abs{ \ket{ \psi_{cl} (t_{i+1}) } }^2$  and 
        $t_{i+1} < t_\text{final}$.
        \item
        If $ r_i \ge \abs{ \ket{ \psi_{cl} (t_{i+1}) } }^2$ and $t_{i+1} \le t_\text{final}$, we determine if a jump occurs.  
        If the simulation parameter \texttt{maxJumps} (see tab. \ref{tab:params}) is larger 
        than the number of previous jumps \texttt{nJumps}, i.e. $\mathtt{nJumps} < \mathtt{maxJumps}$, we perform a jump.
        Otherwise,  in the context of OQS+pNRQCD, we conclude that the subsystem has evolved too far away from the 
        low-lying in-vacuum states to return in a finite amount of time; we thus terminate the quantum trajectory and report total loss for all low-lying bound states (zero overlaps).  In the OQS+pNRQCD framework, a jump as defined by the collapse operators in Eq.~(\ref{eq:collapse}) entails a change of angular momentum, a possible change of color state, and a multiplication of the radial wave-function by $r$. 
        \begin{enumerate} \label{step:general_evolution}
            \item \label{step:angular_mom_jump}
            For $l=0$, the state jumps to $l=1$ with probability 1.
            For $l>0$, we draw a random number $r_{il} \in [0,1]$ to 
            determine whether to jump up or down in the angular momentum quantum number.
            We denote the jumped value of the angular momentum quantum number $l'$ and the trajectory $\ket{ \psi_{cl'}(t_{i+1}) }$.
            \item \label{step:color_jump}
            For $c=s$, i.e. a singlet state, a jump to an octet state occurs with probability 1.
            For $c=o$, i.e. an octet state, we draw an additional random number 
            $r_{ic} \in [0,1]$ to determine whether to jump in the internal quantum number from octet to 
            octet or octet to singlet. 
            We denote the jumped value of the internal quantum number $c'$ and the trajectory $\ket{ \psi_{c'l'}(t_{i+1}) }$.
            \item
            The final form of the jumped state is obtained by multiplying $\ket{ \psi_{c'l'}(t_{i+1}) }$ by the radial part of the 
            jump operator and normalizing, i.e. $\ket{ \tilde{\psi}_{c^\prime l^\prime}(t_{i+1}) } = r \ket{ \psi_{cl}(t_{i+1}) } / ||r \ket{ \psi_{cl}(t_{i+1}) }||$;
            the change in angular momentum and/or internal quantum number is implemented in steps~\ref{step:angular_mom_jump} and~\ref{step:color_jump}. 
        \end{enumerate}
        We proceed to the $(i+1)$\,th iteration of the cycle for this quantum trajectory, i.e. generate a new random number $r_{i} \in [0,1]$ and repeat step~\ref{step:general_evolution} with $\ket{ \tilde{\psi}_{c^\prime l^\prime}(t_{i+1}) }$.
        Note that in the subsequent evolution with $H_{\text{eff}}$ the angular momentum and internal quantum number dependent terms, i.e. the singlet or octet potentials $V_{s,o}(r)$ as defined below Eq.~(\ref{eq:hsoCoul}), are realized using the jumped values of the quantum numbers in the subsequent evolution, i.e. $l'$ and $c'$.
        \item
        Each set of final state overlap ratios or overlaps together with metadata is written to disk as specified in 
        \ref{sec:outputformats}.
    \end{enumerate}
    \item
    The final state overlaps from all quantum trajectories are written to {\tt output/ratios.tsv}. 
    We describe how to process and visualize the results in Section \ref{sec:notebooks}.
\end{enumerate}
Note that the evolution with a phenomenological potential is performed in the same way, 
except that there are no jumps; hence, without jumps, step~\ref{step:jumps} in its entirety in the 
summary of the execution does not exist.

\section{Key parts of the code}
\label{sec:key}

All C source code for the \qtraj package can be found in the {\tt build/src} and {\tt build/include} directories.  
The \texttt{.cpp} files contained in {\tt build/src} are

\listingbox{%
\dirtree{%
.1 build/src.
.2 eigensolver\_arma.cpp.
.2 eigensolver\_mkl.cpp.
.2 initialcondition.cpp.
.2 interpolator.cpp.
.2 outputroutines.cpp.
.2 paramreader.cpp.
.2 potential.cpp.
.2 qtraj.cpp.
.2 trajectory.cpp.
.2 wavefunction.cpp.
}
}

The main entry point is {\tt build/src/qtraj.cpp}.  The main routine found in this file calls routines defined in the other files.  In the following subsections, we present a number of key parts of the code to assist users in understanding the core subroutines which evolve the wave-function between quantum jumps and the central logic used to implement stochastically sampled quantum jumps.  Note that lines that begin with ``.'' indicate portions of the code that have been omitted for readability.

\subsection{Split-step pseudo-spectral wave-function update}

Between quantum jumps, \qtraj uses a split-step pseudo-spectral method to evolve the wave-function as described in Sec.~\ref{sec:splitstep}.  This is implemented by the routine {\tt makeStep} shown below and defined in the file {\tt wavefunction.cpp}.  This routine calls the {\tt DST} function which performs the discrete sine transform of the real and imaginary parts of the wavefunction separately and then sets the double complex wave-function variable to the required DST (in-place transform).  The variables {\tt in} and {\tt out} are buffers used to hold intermediate results.  The variable {\tt spaceKernel} contains $\exp(-i dt \, V_\text{eff}  / 2)$, which are the values of the spatial part of the time evolution operator evaluated at each point on the spatial lattice.  Note that as the potential is time dependent the variable {\tt spaceKernel} must be updated at every time step.  The variable {\tt momKernel} contains $\exp(-i dt \, T_\text{kin})$ which are the values of the momentum-part of the time evolution operator evaluated at each point on the conjugate momentum lattice.  Note that the kinetic energy operator is time-independent, and, hence, {\tt momKernel} only needs to be loaded during the initialization stage.
\begin{lstlisting}
void makeStep(fftw_plan p, dcomp* wfnc, double* in, double* out, dcomp* spaceKernel, dcomp* momKernel)
{
    // make one spatial half-step
    for (int i=0; i<num; i++) wfnc[i] = spaceKernel[i]*wfnc[i];
    // forward DST transform
    DST(p, wfnc, in, out);
    // make one full step in momentum space
    for (int i=0; i<num; i++) wfnc[i] = momKernel[i]*wfnc[i];
    // backward DST transform; it's its own inverse
    DST(p, wfnc, in, out);
    // make one spatial half-step
    for (int i=0; i<num; i++) wfnc[i] = spaceKernel[i]*wfnc[i];
}
\end{lstlisting}

\subsection{Wave-function evolution without jumps}

The \qtraj code includes the capability to run with and without quantum jumps.  If jumps are turned off, then \qtraj simply evolves the wave-function using the complex effective Hamiltonian $H_\text{eff}$.  This is implemented in the routine {\tt evolveWavefunction} listed below.  This routine takes two arguments: {\tt nStart} which is the integer index of the starting time for the evolution and {\tt nSteps} which is the number of time steps to evolve the wave-function forward in time.
\begin{lstlisting}
double evolveWavefunction(int nStart, int nSteps) {
    
    double norm = 1;
    double rexp;
    
    // begin time loop
    for (int n=nStart;n<nStart+nSteps;n++) {
        
        // update the space kernel
        loadSpaceKernel(T[n]);
        
        // output info to screen and/or disk
        if (n%snapFreq==0 || n==nStart) {
            norm = computeNorm(wfnc);
            outputInfo(n,norm,false);
        }
        
        // update the wavefunction
        makeStep(p,wfnc,in,out,spaceKernel,momKernel);
    
    } // end time loop
    
    // output final wavefunction, summary info, and ratios
    norm = computeNorm(wfnc);
    outputInfo(nStart+nSteps,norm,false);
    return norm;
    
}
\end{lstlisting}

\subsection{Wave-function evolution with jumps}

A compactified listing of the {\tt evolveWavefunctionWithJumps} routine found in the file {\tt trajectory.cpp} is shown below (see also Section ~\ref{ssec:algo}).  
The first argument {\tt nStart} is the index of the starting time for the evolution and the second argument {\tt rMin} is the minimum initial random number to consider.  
This variable is computed by first running a complete $H_\text{eff}$ evolution to determine the final norm.
As any random number smaller than the final $H_\text{eff}$ norm results in no quantum jumps, there is no reason to check for jumps.  
In this case, the routine sets the overlaps to those determined from the $H_\text{eff}$ evolution performed during initialization.

In addition, as can be seen from the code listing below, the global parameter {\tt rMax} sets the maximum initial random number accepted by the algorithm.  
If the initial random number is less than {\tt rMax}, \qtraj assumes that the stochastic evolution will result in too many jumps for a significant final overlap with singlet bound states.  
The default for this parameter is 1; reducing from 1 results in faster code execution.  
This faster execution, however, results in a systematic uncertainty (see Fig.~\ref{fig:scaling4} and surrounding discussion).

After this initial check, the evolution proceeds by generating a random number uniformly distributed between 0 and 1, then evolving the wave-function until the norm squared of the wave-function drops below this number.  
At this point, a quantum jump is triggered and two additional random numbers are generated to determine the outgoing color and angular momentum quantum numbers.  
Finally, the wave-function is multiplied by $r$ and re-normalized using the {\tt doJump} subroutine.  
The routine then continues the time loop by generating a new random number and evolving the wave-function forward in time with the updated quantum numbers.  
In all cases, the final result from this routine is stored in a double array called {\tt singletOverlaps}. 
\begin{lstlisting}
void evolveWavefunctionWithJumps(int nStart, double rmin) {
    .
    .
    // generate initial random number
    double r = Random();
    firstRandomNumber = r; // save this for testing output
    // check to see if jumps will be triggered at all
    if (r<rmin) {
        print_line();
        cout << "==> No jumps necessary.  ";
        cout << "Setting overlaps to Heff overlaps." << endl;
        retrieveNoJumpsSingletOverlaps();
        return;
    }
    // check to see if upper threshold is crossed
    if (r > rMax) {
	    print_line();
        cout << "==> Initial random number greater than rmax.  ";
        cout << "Setting overlaps to zero." << endl;
        for (int i=0; i<nBasis; i++) singletOverlaps[i] = 0;
        return;
    }
    .
    .
    // normal evolution until norm drops below random threshold r = Random(0,1) 
    .
    .
    // do the jump if necessary
    if (n<maxSteps) {
        
        // output some info just before the jump
        norm = computeNorm(wfnc);
        outputInfo(n,norm,true);
        
        // determine new angular momentum state
        if (Random() < AngMomProb(lval)) lval -= 1;
        else lval += 1;
        
        // determine new color state
        if (cstate == OCTET  && Random() < 2./7.) cstate = SINGLET;
        else cstate = OCTET;
        
        // apply jump operator and normalize to 1
        doJump(wfnc);
        
        // generate new random # and increment nJumps
        r = Random();
        nJumps++;
        
        // output some info just after the jump
        norm = 1; // forced to be one by doJump so don't waste time computing it
        outputInfo(n,norm,true);
    }
    
    // check to see if we exceeded the maximum number of jumps
    if (nJumps>=maxJumps) {
        print_line();
        cout << "==> Exceeded max jumps.  Terminating evolution." << endl;
        print_line();
        break;
    }
    .
    .
    .    
}

\end{lstlisting}

\begin{figure}[t]
\begin{center}
\fbox{\includegraphics[width=0.85\linewidth]{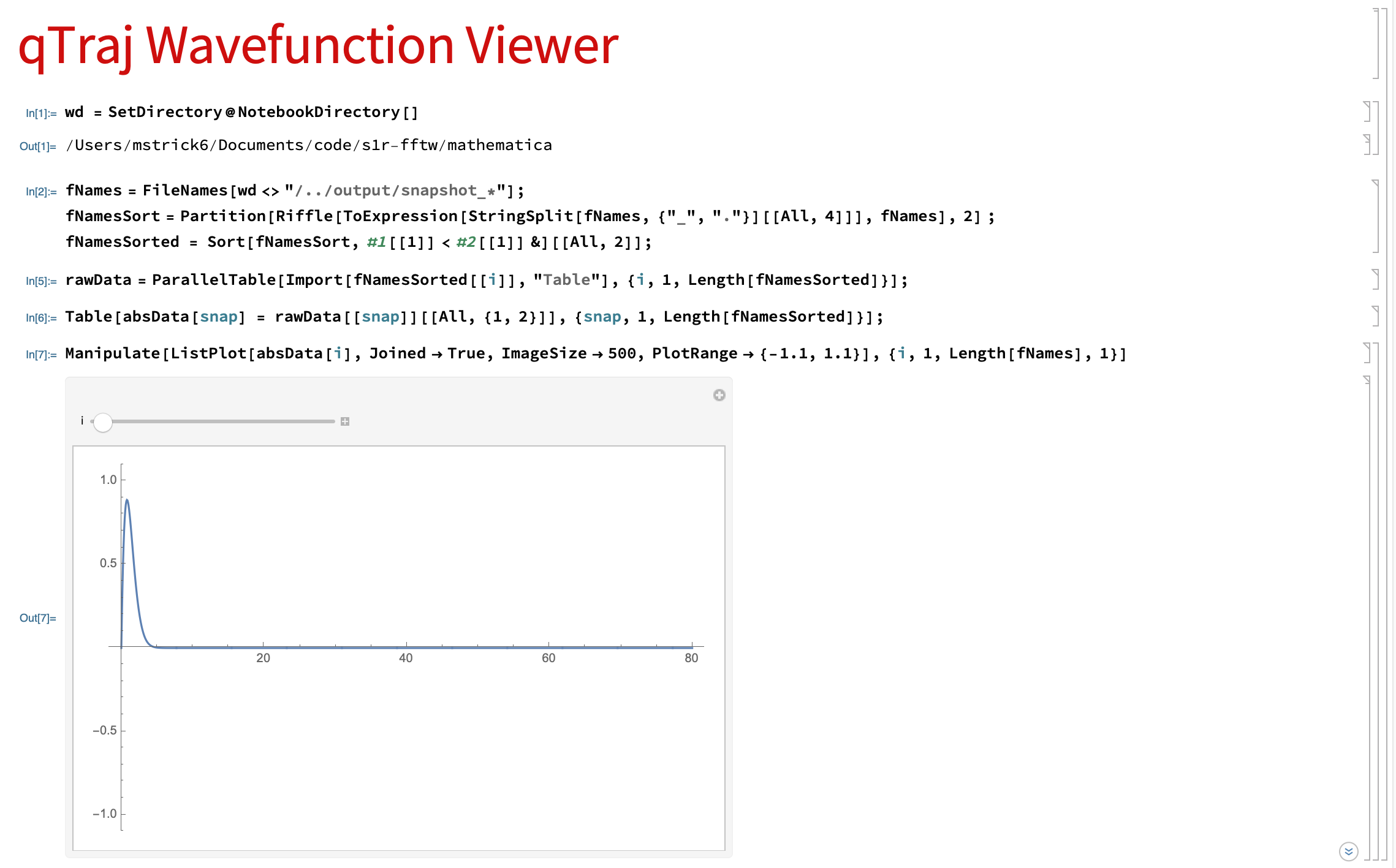}}
\end{center}
\caption{Example {\tt viewer.nb} output.} 
\label{fig:viewer-example}
\end{figure}

\section{Data analysis notebooks}
\label{sec:notebooks}

\qtraj can be set to output snapshots of the evolution of the wave-function in the directory {\tt output} or {\tt output-<seed>} depending on the value of {\tt dirnameWithSeed} (see Sec.~\ref{sec:params}).
The format for these files is detailed in \ref{sec:outputformats}.
The user can suppress the output of these snapshots by setting {\tt saveWavefunctions} to 0.  The only other disk output from \qtraj is to the file {\tt ratios.tsv} in the output directory.  
The format of this file depends on the {\tt temperatureEvolution} parameter (see Sec.~\ref{sec:params}) as described in~\ref{sec:outputformats}.  
In order to assist the user in understanding how to process the resulting output, we include Mathematica notebooks that read all three output formats.

The {\tt mathematica} directory in the main package contains Mathematica \cite{Mathematica} notebooks that can be used for data analysis:
\listingbox{%
\dirtree{%
.1 mathematica.
.2 raaCalculator.nb.
.2 raaCalculator-trajectories.nb.
.2 ratioFileReader.nb.
.2 viewer.nb.
}
}

\subsection{The notebook {\tt viewer.nb}}

The Mathematica notebook {\tt viewer.nb} allows the user to visualize the wave-function evolution produced by \qtraj.  
To generate the files necessary, the parameter {\tt saveWavefunctions} must be set to 1.
In this configuration, the initial wave-function, ``snapshots'' of the wave-function recorded every {\tt snapFreq} steps, and the final wave-function are exported to the {\tt output} folder as separate files named {\tt snapshot\_\#.tsv} where \# is the step in the evolution.  

To generate output for testing, the user can execute the following example
\listingbox{%
\tt
\$ ./qtraj -initType 0 -doJumps 0 -saveWavefunctions 1 -dirnameWithSeed 0
}
Upon successful completion, one can evaluate all cells in the {\tt viewer.nb} notebook.
Typical output is shown in Fig.~\ref{fig:viewer-example}.

\subsection{The notebook \texorpdfstring{{\tt ratioFileReader.nb}}{ratioFileReader.nb}}

\begin{figure}[ht]
\begin{center}
\fbox{\includegraphics[width=0.8\linewidth]{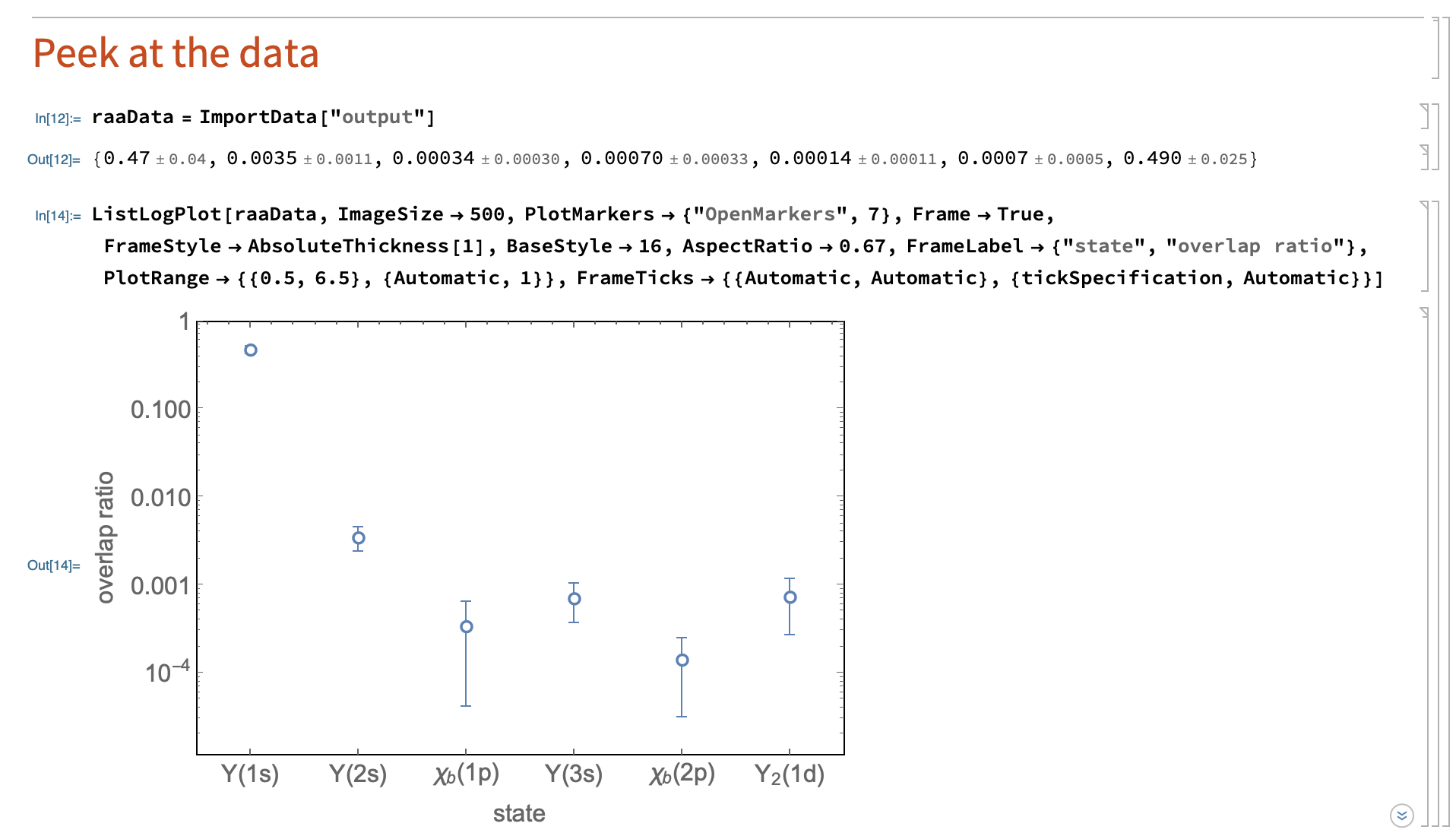}}
\end{center}
\caption{Example {\tt ratioFileReader.nb} output.} 
\label{fig:ratioFileReader-example}
\end{figure}

To generate output for testing, the user can execute
\listingbox{%
\tt
\$ ./qtraj -initType 0 -doJumps 1 -nTrajectories 128 -dirnameWithSeed 0
}
Upon successful completion, the code generates a file {\tt output/ratios.tsv} which contains the resulting survival probabilities (or overlaps) for the $\Upsilon(1S),~\Upsilon(2S),~\chi_b(1P),~\Upsilon(3S),~\chi_b(2P),~\text{and}~\Upsilon_2(1D)$ states.  
To visualize the data, the user can open the {\tt ratioFileReader.nb} Mathematica notebook and evaluate all cells in the notebook.  
The first section of the notebook produces data for the ratios averaged over all trajectories based on the output generated in {\tt output/ratios.tsv}. 
A summary plot similar to Fig.~\ref{fig:ratioFileReader-example} is generated.  
To add more statistics, the user can rerun the command above.  
The newly generated results are appended to {\tt output/ratios.tsv}.

\subsection{The notebook {\tt raaCalculator-trajectories.nb}}

Note that, although the example above runs on a single core, access to a computing facility supporting Portable Batch Scripts (PBS) enables the user to launch many \qtraj trajectories simultaneously.\footnote{We also provide two condor scripts which can be used for launching trajectories on the Open Science Grid \cite{osg07,osg09}.}  
For this purpose, we have provided a PBS script {\tt scripts/manyTrajectories.pbs} which automates the extraction of individual physical trajectories from a tgz file containing a large number of Monte-Carlo generated physical trajectory files individually having the format specified in~\ref{sec:formats2}.  The tgz file should be saved in the following location: {\tt input/runset/trajectories.tgz}.  
An example trajectories.tgz file is available in~\cite{kent-code-library}. 

The {\tt manyTrajectories.pbs} script has adjustable variables at the top allowing the user to set the number of cores on which to launch jobs and the number of physical and quantum trajectories to generate.

\begin{lstlisting}[language=Bash]
#PBS -l walltime=8:00:00
#PBS -l nodes=1:ppn=1
#PBS -l mem=300MB
#PBS -t 1-100
#PBS -N qtraj-trajectory
#PBS -j oe
#PBS -A PGS0253

module load fftw3

# number of physical trajectories to sample
NPTRAJ=100

# number of quantum trajectories to sample per physical trajectory
NQTRAJ=50
  .
  .
  .
# extract random unique set of NPTRAJ trajectories from trajectories.tgz
tar -ztf $PBS_O_WORKDIR/input/runset/trajectories.tgz | shuf -n ${NPTRAJ} > trajectoryList.txt
tar -xzf $PBS_O_WORKDIR/input/runset/trajectories.tgz --warning=no-timestamp --files-from trajectoryList.txt
  .
  .
  .
for file in ./input/runset/*
do
    echo "Running with $file"
    # run the code first for l=0 then for l=1
    ./qtraj -initL 0 -nTrajectories ${NQTRAJ} -temperatureEvolution 2 -temperatureFile $file > log_0_${cnt}.txt 2> err_0_${cnt}.txt
    ./qtraj -initL 1 -nTrajectories ${NQTRAJ} -temperatureEvolution 2 -temperatureFile $file > log_1_${cnt}.txt 2> err_1_${cnt}.txt
done
  .
  .
  .
\end{lstlisting}

\begin{figure}[t]
\begin{center}
\fbox{\includegraphics[width=0.9\linewidth]{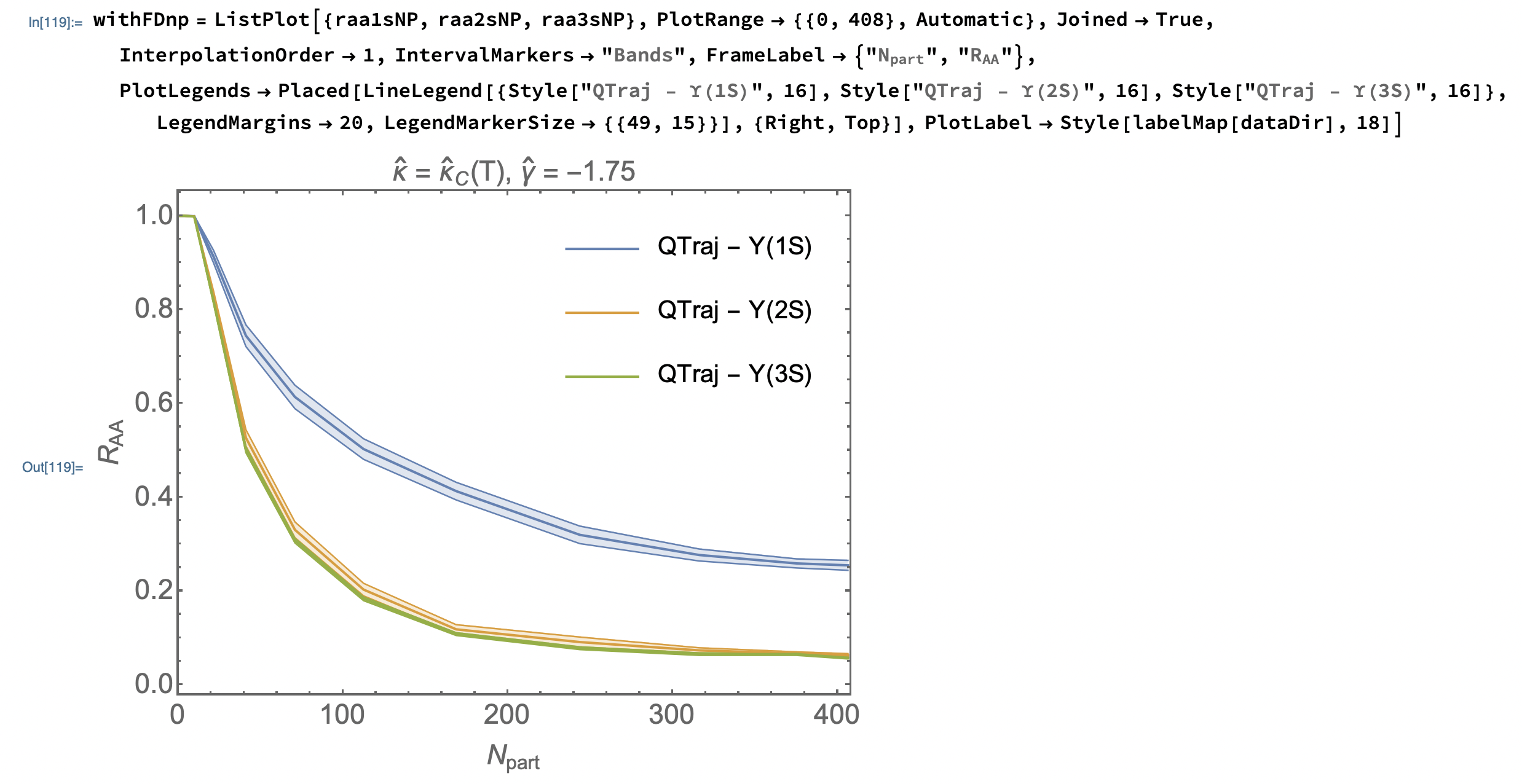}}
\end{center}
\caption{Example {\tt raaCalculator-trajectories.nb} output.} 
\label{fig:raaCalculator-trajectories}
\end{figure}

For post-processing, all {\tt ratio.tsv} files produced during execution of the {\tt manyTrajectories.pbs} script should be concatenated and combined into a single gzipped file, canonically named {\tt datafile.gz}.  This can be done using the UNIX {\tt cat} command.  This step is the responsibility of the user since different computing environments have potentially different formats for the output directory structure, etc.  To process the data, the user should create a new directory in the {\tt input} subfolder of the {\tt mathematica} folder and place the {\tt datafile.gz} in the new folder.  An example {\tt datafile.gz} file is supplied with the \qtraj distribution and can be found in the {\tt mathematica/input/example1} folder.  The supplied version of {\tt mathematica/raaCalculator-trajectories.nb} is setup to read in this example and, upon successful execution, generate a plot similar to that shown in Fig.~\ref{fig:raaCalculator-trajectories}.

\subsection{The notebook {\tt raaCalculator.nb}}

To make runs using physical-trajectory averaged temperatures files, the \qtraj package includes a PBS script called {\tt scripts/scanImpactParameters.pbs}.  The header and topmost variables of this script allow the user to specify the number of cores on which to run the code and the number of trajectories to generate per core.

\begin{lstlisting}[language=Bash]
#PBS -l walltime=08:00:00
#PBS -l nodes=1:ppn=1
#PBS -l mem=300mb
#PBS -t 1-100
#PBS -N qtraj-fftw
#PBS -j oe
#PBS -A PGS0253

module load fftw3

# number of quantum trajectories per impact parameter for each job
# total number of trajectories is the size of job array times this number
NTRAJ=2048

# impact parameter list
blist=("0" "2.32326" "4.24791" "6.00746" "7.77937" "9.21228" "10.4493"\
"11.5541" "12.5619" "13.4945" "14.3815")

  .
  .
  .
\end{lstlisting}

The example above launches \qtraj on 100 cores, with each instance generating 2048 quantum trajectories.  Once combined, this results in 204,800 quantum trajectories per point in {\tt blist}.
It is the user's responsibility to adjust the number of requested nodes using the line {\tt PBS -t 1-100}. The range specified indicates the number of independent machines (cores) to use.  The amount of time specified by {\tt PBS -l walltime=02:00:00} is the expected maximum time for each core to complete the task.  The user can set the number of quantum trajectories for each core to simulate using the variable {\tt NTRAJ}.  
Finally, there is a list of impact parameters over which to loop.  
These should map to the input files of the form {\tt input/temperature/averageTemperature\_<impact parameter>.txt}.\footnote{A set of such files is included in the standard distribution.}

After adjusting {\tt scripts/scanImpactParameters.pbs} to their needs, from the main directory the user can execute
\listingbox{%
\tt
\$ qsub ./scripts/scanImpactParameters.pbs
}
to launch a job.
Upon successful completion, a directory, or directories depending on the user's environment, of the form {\tt output-$<$jobid$>$} are created.  These each contain sub-folders for every impact parameter in {\tt blist}.  
Inside of each impact parameter's folder, there is a file named {\tt ratios.gz}, which is a compressed collection of {\tt ratio.tsv} style outputs.  
If the user's system generates multiple output folders, the user will need to combine the ratios.gz files for each impact parameter.  
This can be done using the {\tt cat} utility in UNIX, e.g., {\tt cat output-*/0/ratios.gz > ./combined/0/ratios.gz}.
We provide a shell script {\tt scripts/combine.sh} to facilitate this process.

To plot the computed survival probabilities, the user can open the Mathematica notebook {\tt raaCalculator.nb} and adjust the paths to point to the local output directory (or the combined one created using {\tt cat} or {\tt scripts/combine.sh}).  
Upon successful execution, the user is presented with a plot similar to the one shown in Fig.~\ref{fig:raaCalculator-trajectories}. We note that, although the name of the notebook is {\tt raaCalculator.nb}, this example notebook computes only the pre-feeddown survival probabilities and not $R_{AA}$ which requires further processing by the user.
We provide sample data with the \qtraj distribution called {\tt mathematica/input/example2}. 
The Mathematica notebook {\tt raaCalculator.nb} is configured to import data from this folder as an example for the user.

\begin{figure}[t]
\begin{center}
\fbox{\includegraphics[width=0.9\linewidth]{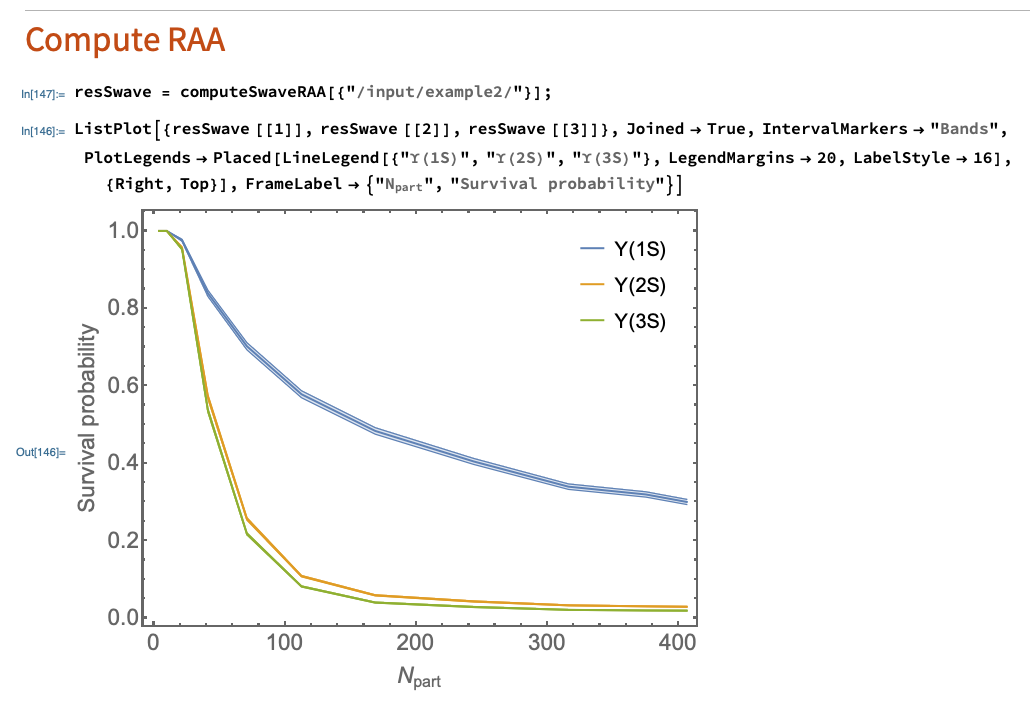}}
\end{center}
\caption{Example {\tt raaCalculator.nb} output.} 
\label{fig:raaCalculator}
\end{figure}

\section{Code performance and scaling tests}
\label{sec:tests}

The \qtraj distribution includes a suite of built-in self tests that can be run using {\tt make~tests}.  This functionality requires that the Google testing framework be installed (see \ref{subsec:installgtest}).  In addition, we note that apples-to-apples comparisons of \qtraj and a QuTiP 2 implementation of the Lindblad equation solver were performed in~\cite{Brambilla:2020qwo}.  Therein it was demonstrated that \qtraj and QuTiP 2 implementations of the Lindblad equation solver are in agreement when the same lattice sizes, angular momentum cutoff, etc. are used.

In this section, we present the dependence of \qtraj results on various run time parameters, e.g. lattice spacing, lattice volume, temporal step size, maximum number of jumps, etc.  For all results presented in this section, we use Gaussian S-wave initial conditions with $\Delta = 0.2 a_0$ and Bjorken temperature evolution starting at $\tau = 0.6$ fm with $T_0=$ 425 MeV and terminating at $T_f = 250$ MeV.  

\begin{figure}[t]
\begin{center}
\includegraphics[width=0.45\linewidth]{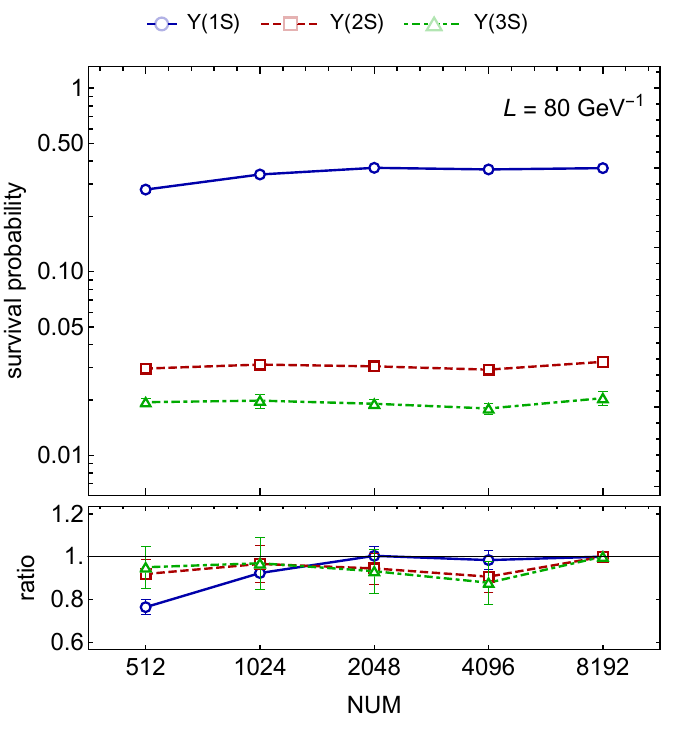} \hspace{5mm} 
\includegraphics[width=0.45\linewidth]{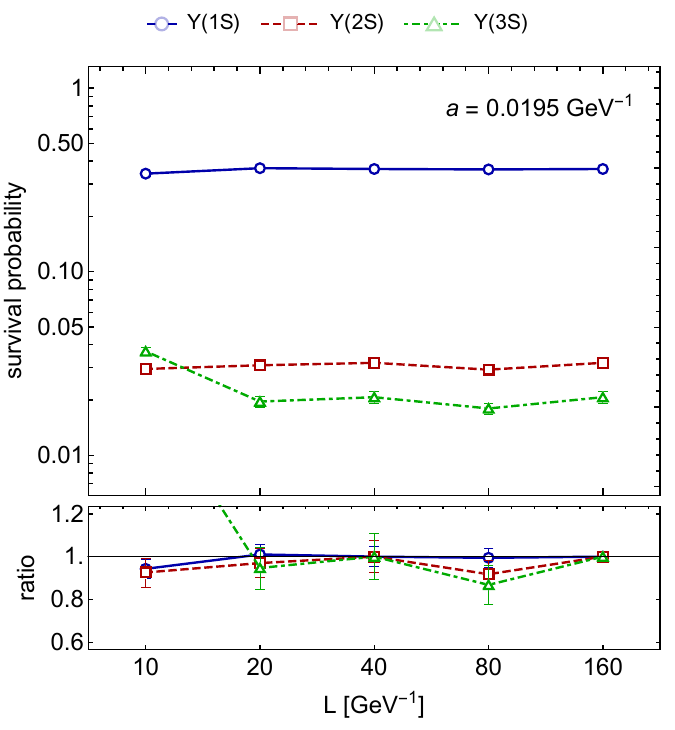}
\end{center}
\caption{(Left) Survival probabilities for $\Upsilon(1S)$, $\Upsilon(2S)$, and $\Upsilon(3S)$ versus the number of lattice sites for a fixed volume $L = 80\ \text{GeV}^{-1}$.  (Right) Same versus the lattice volume $L$ for fixed $a = 0.0195 \text{GeV}^{-1}$.  For both panels, we used Gaussian S-wave initial conditions with $\Delta = 0.2 a_0$ and Bjorken temperature evolution starting at $\tau = 0.6$ fm with $T_0=$ 425 MeV.  Error bars indicate the statistical uncertainty associated with averaging over 100,000 quantum trajectories.} 
\label{fig:scaling}
\end{figure}

In the left panel of Fig.~\ref{fig:scaling}, we present the dependence of the extracted $\Upsilon(1S)$, $\Upsilon(2S)$, and $\Upsilon(3S)$ survival probabilities on the lattice spacing with a fixed lattice volume of $L = 80\ \text{GeV}^{-1}$.  We obtain the survival probabilities by computing the ratio of the modulus squared of each eigenstate's overlaps with the final and initial wave-functions. We note that the square of an eigenstate overlap with the initial wave-function corresponds, in our model, to the probability of measuring that state if there were no medium. This ratio is averaged over 100,000 quantum trajectories and we report the mean along with the uncertainties obtained by computing the standard error of the mean.  As can be seen from the left panel of Fig.~\ref{fig:scaling}, there is only a weak dependence on the lattice spacing.  In order to quantify this variation, we note that the ratios of the $\Upsilon(1S)$, $\Upsilon(2S)$, and $\Upsilon(3S)$ survival probabilities obtained using 8192 and 4096 lattice points is $\{1.02 \pm 0.04,1.11 \pm 0.08,1.15 \pm0.11\}$, respectively. Note that the effects of the radial discretization are naturally expected to be more severe for states with larger contributions from small radii. For past phenomenological studies, 4096 points were used with a volume of $L = 80\ \text{GeV}^{-1}$ \cite{Brambilla:2020qwo}. 

In the right panel of Fig.~\ref{fig:scaling}, we present the dependence of the extracted $\Upsilon(1S)$, $\Upsilon(2S)$, and $\Upsilon(3S)$ survival probabilities on the lattice volume $L$ holding the lattice spacing fixed at $a = 5/256 \ \text{GeV}^{-1}$.  As this figure demonstrates, for the type of initial conditions used, one finds that there is no significant volume effect for $L > 20 \ \text{GeV}^{-1}$.  Again, in order to quantify this variation, we note that the ratios of the $\Upsilon(1S)$, $\Upsilon(2S)$, and $\Upsilon(3S)$ survival probabilities obtained using $L=80$ and 160 $\text{GeV}^{-1}$ are $\{1.01\pm0.04,1.10\pm0.07,1.16\pm0.10\}$, respectively.  Taken together, these results demonstrate that with 100,000 trajectories the systematic uncertainties associated with finite lattice size and spacing are on the order of 5\%, 10\%, and 15\% for the $\Upsilon(1S)$, $\Upsilon(2S)$, and $\Upsilon(3S)$ survival probabilities, respectively. Note that the effects of the finite volume are naturally expected to be more severe for states with larger contributions from large radii.

\begin{figure}[t]
\begin{center}
\includegraphics[width=0.45\linewidth]{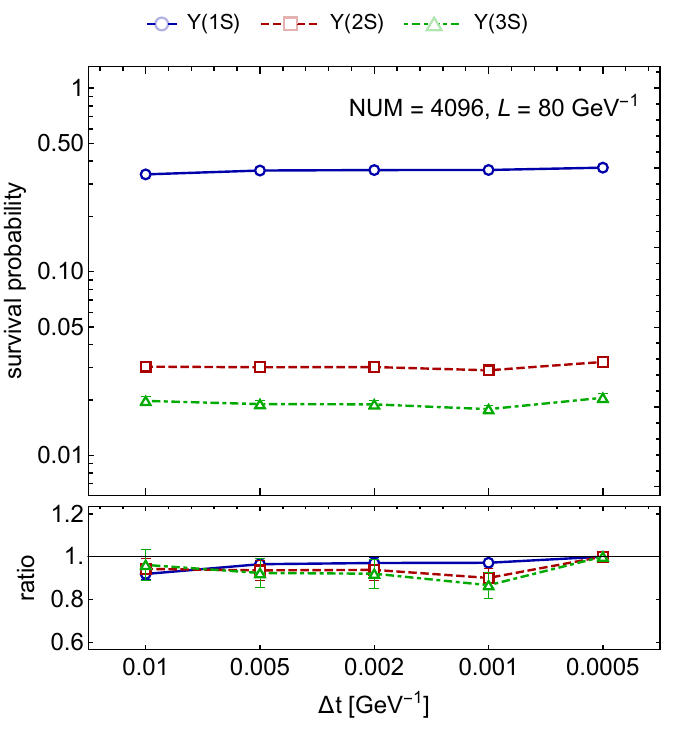} \hspace{5mm}
\includegraphics[width=0.45\linewidth]{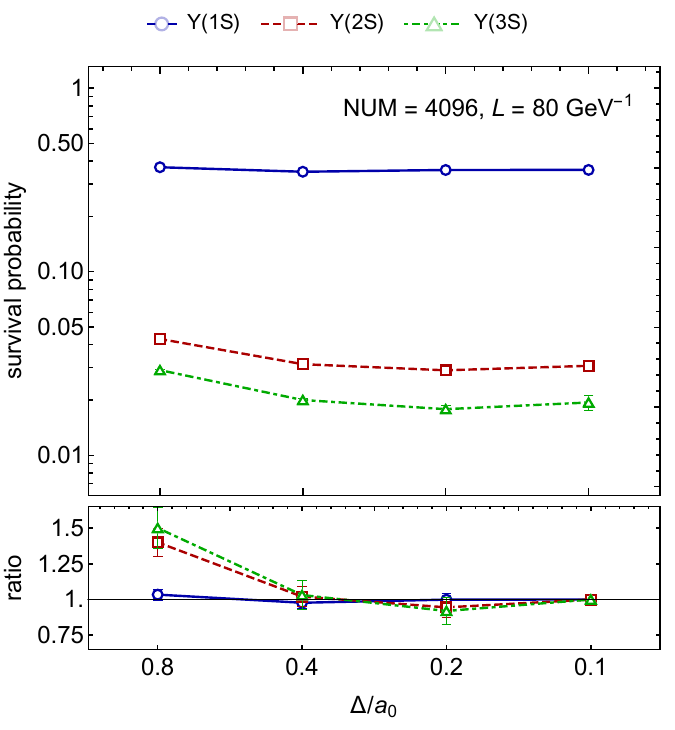}
\end{center}
\caption{(Left) Survival probabilities for $\Upsilon(1S)$, $\Upsilon(2S)$, and $\Upsilon(3S)$ versus the time interval $\Delta t$ for a fixed volume $L = 80\; \text{GeV}^{-1}$ and $NUM=4096$.  (Right) Same as a function of the width of the initial Gaussian $\Delta/a_0$.  For the left panel we used Gaussian S-wave initial conditions with $\Delta = 0.2 a_0$.  For both panels, we used Bjorken temperature evolution starting at $\tau = 0.6$ fm with $T_0=$ 425 MeV.  In both panels, the error bars indicate the statistical uncertainty associated with averaging over quantum trajectories.} 
\label{fig:scaling2}
\end{figure}

Turning next to the dependence of the survival probabilities on the temporal lattice spacing $\Delta t$ ({\tt dt}), in the left panel of Fig.~\ref{fig:scaling2}, we present the $\Upsilon(1S)$, $\Upsilon(2S)$, and $\Upsilon(3S)$ survival probabilities as a function of $\Delta t$.  For each $\Delta t$, we keep $L = 80\ \text{GeV}^{-1}$ and $\texttt{NUM}=4096$ and average over 100,000 quantum trajectories.  As can be seen from this figure, the results show a weak dependence on the assumed temporal lattice spacing.  In this case, we find that the ratios of the $\Upsilon(1S)$, $\Upsilon(2S)$, and $\Upsilon(3S)$ survival probabilities obtained using $\Delta t=0.005$ and 0.001 $\text{GeV}^{-1}$ are $\{1.029\pm0.03,1.11\pm0.05,1.15\pm0.08\}$, respectively.  From this, we can estimate the uncertainties associated with the temporal lattice spacing to be on the order of 5\%, 10\%, and 15\% for the $\Upsilon(1S)$, $\Upsilon(2S)$, and $\Upsilon(3S)$ survival probabilities, respectively.

In the right panel of Fig.~\ref{fig:scaling2}, we present the dependence of the extracted 1S, 2S, and 3S survival probabilities on the initial width $\Delta$ of the Gaussian wave-function used.  As can be seen from this figure, the 1S survival probability does not depend strongly on the assumed value of $\Delta$, whereas the excited states survival probablitities only become approximately independent of $\Delta$ for $\Delta \lesssim 0.2$.  In order to quantify the change, we note that the ratios of the $\Upsilon(1S)$, $\Upsilon(2S)$, and $\Upsilon(3S)$ survival probabilities obtained using $\Delta/a_0=0.1$ and 0.02 are $\{1.00\pm 0.04,1.06\pm 0.08,1.09\pm 0.11\}$, respectively.  These are all consistent with unity within statistical errors; however, there is an indication that the corrections are larger for excited states.  We note that when using $\Delta/a_0 = 0.1$ we doubled the number of quantum trajectories in order to obtain similar uncertainty estimates as $\Delta/a_0 = 0.2$ due to the fact the system jumps more frequently for smaller $\Delta$.  For this reason, for phenomenological applications, we use $\Delta/a_0 = 0.2$ since it requires substantially less computational effort.

\begin{figure}[t]
\begin{center}
\includegraphics[width=0.45\linewidth]{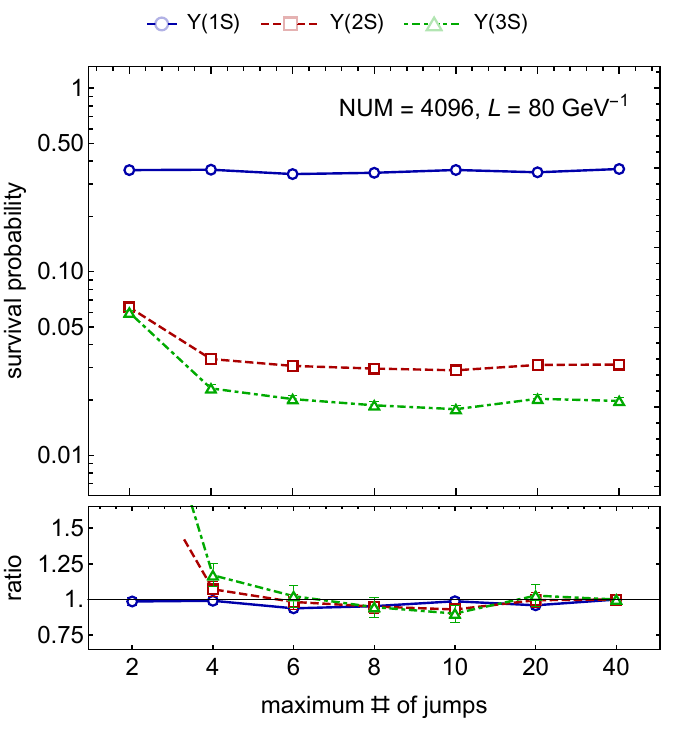} \hspace{5mm}
\includegraphics[width=0.45\linewidth]{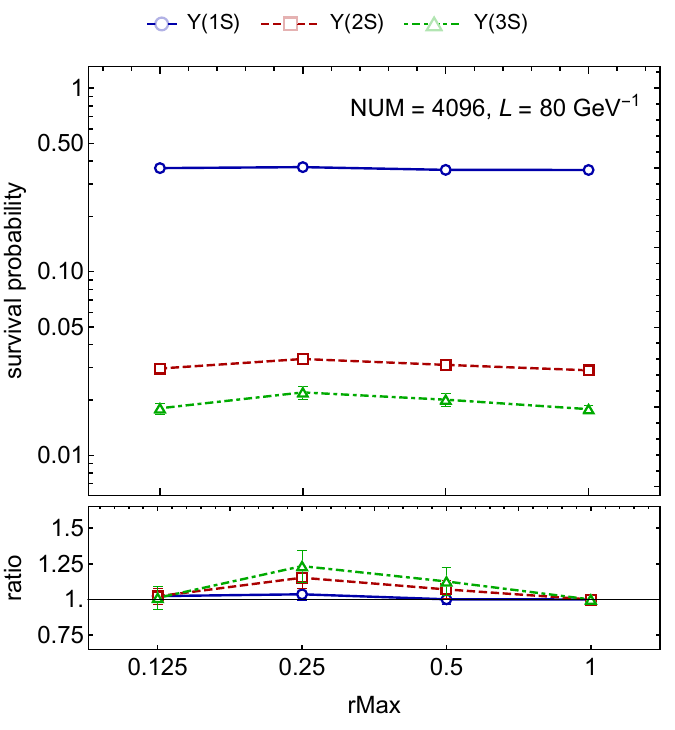}
\end{center}
\caption{(Left) Survival probabilities for $\Upsilon(1S)$, $\Upsilon(2S)$, and $\Upsilon(3S)$ versus the maximum number of jumps allowed for a fixed volume $L = 80\ \text{GeV}^{-1}$ and $\text{NUM}=4096$.  (Right) Same versus {\tt rMax}.  For both panels, we used Gaussian S-wave initial conditions with $\Delta = 0.2 a_0$ and Bjorken temperature evolution starting at $\tau = 0.6$ fm with $T_0=$ 425 MeV.  Error bars indicate the statistical uncertainty associated with averaging over 100,000 quantum trajectories.} 
\label{fig:scaling3}
\end{figure}

Finally, we investigate the dependence of \qtraj results on two optimization parameters: {\tt maxJumps} and {\tt rMax}.  The former sets the maximum number of jumps before the state evolution is terminated and the latter sets the maximum value for the initial random number used during quantum trajectory evolution.  We generate the initial random number in the range $(0,1)$; however, if the generated initial random number is greater than {\tt rMax}, we set the survival probability for the quantum trajectory sampled to zero.  This is justified by the fact that trajectories that jump early in the quantum trajectory evolution will have essentially zero overlap with the singlet bound states of interest.  In the left panel of Fig.~\ref{fig:scaling3}, we present the 1S, 2S, and 3S survival probabilities as a function of ${\tt maxJumps}$.  As this panel demonstrates, the results for the $\Upsilon(1S)$ survival probability do not depend strongly on ${\tt maxJumps}$, whereas the two excited states have more dependence on this parameter. 
Turning to the right panel of Fig.~\ref{fig:scaling3}, we present the 1S, 2S, and 3S survival probabilities as a function of ${\tt rMax}$.  From this figure, we once again see that the 1S state survival probability is independent of this parameter while the excited states survival probabilities have a weak dependence on ${\tt rMax}$.  In practice, ${\tt rMax} < 1$ can be used to accelerate code execution with the speed gain proportional to 1/{\tt rMax}. This functionality is useful for obtaining quick estimates of survival probabilities.

\section{Code benchmarks and comparisons with existing frameworks}
\label{sec:benchmarks}

A key benefit of \qtraj compared to methods relying on direct solution of the reduced density matrix evolution is the scaling of the computational complexity like ${\tt NUM} \log{\tt NUM}$, the dominant contribution to which comes from the forward and inverse Fourier Sine transforms.
Methods which solve the matrix evolution equation directly can be expected to scale like ${\tt NUM}^\alpha$ with $\alpha \gtrsim 3$.  Additionally, in terms of memory use, the memory footprint of \qtraj scales like ${\tt NUM}$, whereas in this respect matrix-based methods scale like ${\tt NUM}^2$.  Since, in practice, one must have large spatial grids in order to properly resolve both bound and unbound state dynamics, this places a strong limitation on such methods.  We also note that since each quantum trajectory sampled by \qtraj is statistically independent, one can parallelize the production of \qtraj results straightforwardly.

\begin{figure}[t]
\begin{center}
\includegraphics[width=0.44\linewidth]{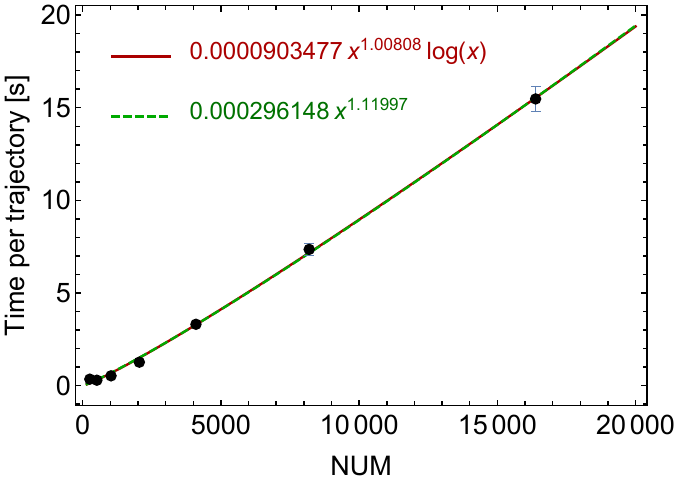} \hspace{5mm}
\includegraphics[width=0.44\linewidth]{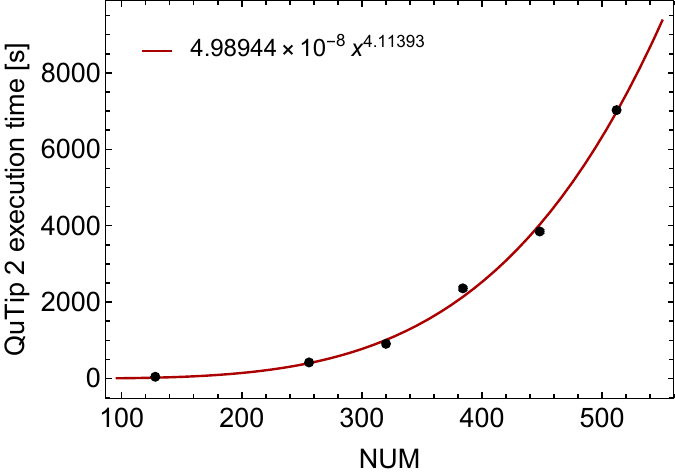}
\end{center}
\caption{(Left) Time in seconds required to compute one quantum trajectory using \qtraj as a function of the number of discrete points in the lattice, {\tt NUM}.  Data were obtained from average using 5 runs for each {\tt NUM} which generate 100 quantum trajectories.  For the measurement, we used the UNIX utility {\tt perf}.  Solid and dashed lines show the result of two fits to these data.  (Right) Time in seconds to compute the evolution of the density matrix with $l_{\rm max} = 1$ using QuTiP 2 as a function of the {\tt NUM}.  Data were again obtained by averaging over 5 runs for each {\tt NUM}.  Solid line shows a fit to the collected data.  In both left and right panels Gaussian S-wave initial conditions were used.} 
\label{fig:performance}
\end{figure}

In order to compare the two methods, we use the code developed in~\cite{Brambilla:2016wgg,Brambilla:2017zei} which solves the same problem, but uses the \textit{mesolve} function of the popular QuTiP 2 Python package for describing the dynamics of open quantum systems \cite{qutip2,Brambilla:2016wgg,Brambilla:2017zei}.  The same initial conditions, namely singlet S-wave Gaussian initial conditions with $\Delta=0.2$ are considered.  In the left panel of Fig.~\ref{fig:performance}, we present the number of seconds required per trajectory as a function of the lattice size ${\tt NUM}$.  For this measurement, we use the UNIX {\tt perf} utility and measured the mean run-time per trajectory along with the standard error of the mean. All tests were performed on an a single Intel Xeon E5-2630 v3 CPU, which runs at a clock rate of 2.40 GHz.  The total memory available to the runs was 64 GB and the operating system used was Red Hat Enterprise Linux Server release 7.9.

The solid and dashed lines in Fig.~\ref{fig:performance} show the result of two fits to these data.  The solid red line indicates the result obtained when assuming a fit function of the form $A \, {\tt NUM}^B \, \log{\tt NUM}$.  We note, importantly, that the extracted coefficient $B$ is very close to unity in this case, suggesting that the scaling of the \qtraj execution time is dominated by the underlying FFTs which is predicted to give scaling of the form ${\tt NUM} \log{\tt NUM}$.  As an alternative, the green dashed line shows results obtained using a fit form $A \, {\tt NUM}^B$.  As one can see, in this range the two fits to the \qtraj performance data are virtually indistinguishable and one finds that they have similar $\chi^2$ per degree of freedom.  We find that these two fits to the \qtraj single trajectory time are within 5\% of one another for $512 \leq {\tt NUM} \leq 2.5 \times 10^5$.

In the right panel of Fig.~\ref{fig:performance}, we present run-times for the QuTiP 2 based solver \cite{qutip2,Brambilla:2016wgg,Brambilla:2017zei}.  In this case, one is restricted to much smaller lattices due to the memory footprint of the matrix-based solvers; ${\tt NUM} = 512$ is the largest lattice that we can run given the total memory of the machine used for testing (64 GB).  From the right panel of Fig.~\ref{fig:performance}, we see that the QuTiP 2 run-time data collected are well-described by a fit of the form $A \, {\tt NUM}^B$ with $B \sim 4$.  
We note that the measurements presented in the left and right panels of Fig.~\ref{fig:performance} are not directly comparable since the left panel presents the single trajectory run times.  In order to compare the total run times, one should multiply the \qtraj single-trajectory estimates by the number of quantum trajectories that are necessary to obtain convergence within a given bound.  In practice, for the application to bottomonium suppression ($R_{AA}$) one finds that, for a single physical trajectory through the quark-gluon plasma, on the order of 100,000 trajectories are required to obtain sufficient statistics (see e.g. Figs.~\ref{fig:scaling}-\ref{fig:scaling3} where typical statistical uncertainties are reported).  Finally, we note that one major advantage of the \qtraj solver is that in this case one does not have to impose a cutoff on the orbital angular momentum $l$, whereas in the QuTiP 2 based implementation an angular momentum cutoff of $l_\text{max}=1$ was used.  If one increases the angular momentum cutoff in the QuTiP 2 based code, the resulting run times become longer and memory limitations more severe, since the dimension of the reduced density matrix scales like $(2 \, {\tt NUM} \, (l_\text{max} + 1))^2$, where the factor of 2 accounts for the color singlet and octet states.  

Based on the performance data collected and discussions above, one can estimate
\be
T_\text{qtraj} \sim \frac{N_\text{traj} \, {\tt NUM} \, \log  {\tt NUM}}{N_\text{cores}} \, ,
\ee
where $N_\text{cores}$ is the number of independent cores available and
\be
T_\text{QuTiP 2}^\text{single\ core} \sim (l_\text{max}+1)^4 \, {\tt NUM}^4 \, .
\ee
As a case study, one can consider the case of ${\tt NUM} =4096$ which was used in the phenomenological predictions contained in~\cite{Brambilla:2020qwo}. Assuming $N_\text{traj} = 10^5$ and $l_\text{max} = 1$, for a single core, one obtains $T_\text{qtraj}^\text{single\ core} \simeq 3 \times 10^5$ seconds, whereas the extrapolated prediction for $T_\text{QuTiP 2}^\text{single\ core}$ is $3.6 \times 10^7$ seconds.  Based on our fits, when $l_\text{max}=1$ and $N_\text{traj}=10^5$, the two codes become comparable in run time for lattices of size ${\tt NUM}_\text{cross}^{l_\text{max}=1} \sim 850$.  For $l_\text{max} \geq 1$, one can expect this point to be at ${\tt NUM}_\text{cross} \sim 1700/(l_\text{max}+1)$, which goes to zero as $l_\text{max}$ becomes large.  Summarizing, we find that for ${\tt NUM} \gtrsim {\tt NUM}_\text{cross}$ \qtraj running on a single core will be more computationally efficient than the QuTiP 2 implementation used in \cite{Brambilla:2016wgg,Brambilla:2017zei}.

\begin{figure}[t]
\begin{center}
\includegraphics[width=0.45\linewidth]{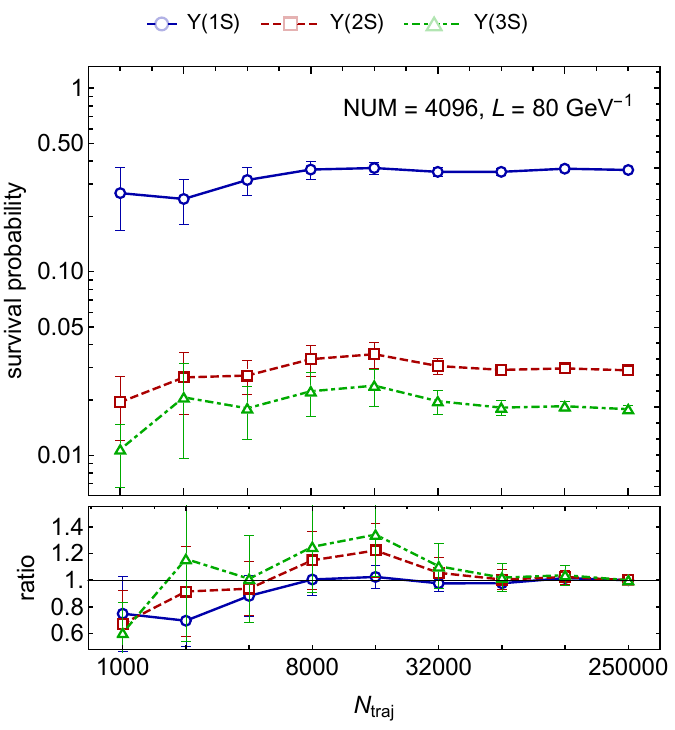} \hspace{5mm}
\includegraphics[width=0.45\linewidth]{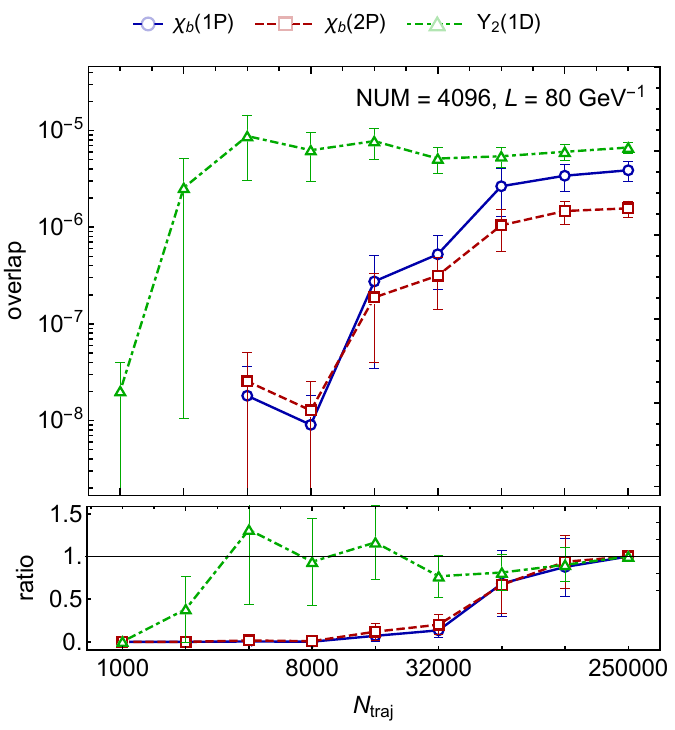}
\end{center}
\caption{(Left) Survival probabilities for $\Upsilon(1S)$, $\Upsilon(2S)$, and $\Upsilon(3S)$ versus the number of quantum trajectories for a fixed volume $L = 80\ \text{GeV}^{-1}$ and $\text{NUM}=4096$.  (Right) Off-diagonal overlaps with the $\chi_b(1P)$, $\chi_b(2P)$, and $\Upsilon(1D)$ states.  For both panels, we used Gaussian S-wave initial conditions with $\Delta = 0.2 a_0$ and Bjorken temperature evolution starting at $\tau = 0.6$ fm with $T_0=$ 425 MeV.  Error bars indicate the statistical uncertainty associated with averaging the considered subset of quantum trajectories.
} 
\label{fig:scaling4}
\end{figure}

Finally, we emphasize that the above comparison is of the single core run time.  Due to the embarrassingly parallel nature of the \qtraj algorithm, one can simply run the code on many distributed cores simultaneously and no communication is required between cores.  In principle, it is also possible to parallelize the QuTiP 2 {\tt mesolve} based code; however, due to the fact that the computation does not easily factorize into independent processes, one can expect significant communication bottlenecks. Based on our results and the discussion above, our conclusion is that, when simulating large lattices, \qtraj provides a superior implementation of the Lindblad equation solver, and it does so without imposing an angular momentum cutoff.  We additionally note that, due to the portable and lightweight nature of the \qtraj code, one can leverage state, national, and international computing facilities, where typically tens of thousands of cores are available, thus greatly accelerating the computation.

Eventually, one must consider how many trajectories are necessary to achieve a given target accuracy. 
We show results using different subsets of the same statistical ensemble of trajectories in Fig. \ref{fig:scaling4} presenting $S$-wave overlaps in the left panel and off-diagonal overlaps in the right panel.
The 1S survival probability is obtained to better than 10\% accuracy already with approximately $10^4$ trajectories, but the higher $S$-waves 
require at least a factor five more statistics and off-diagonal overlaps in general substantially more. 
Accessing the singlet $P$-wave from an initial singlet $S$-wave cannot be accomplished with fewer than three 
jumps; hence, it requires at least two periods of evolution as an octet state between the jumps.
For this reason, substantially more statistics are needed, and the final plateau may not be reached even with 
$2.5\times 10^5$ trajectories, although the convergence is visible in the plot. 
The $D$-wave, which cannot be reached with fewer than two jumps, still appears to converge at a similar 
rate as the excited $S$-waves. 
Although these off-diagonal overlaps have substantially larger relative uncertainties than the $S$-waves, they are 
three orders of magnitude suppressed, rendering them quantitatively irrelevant.

\section{Conclusions}
\label{sec:conclusions}

In this paper, we introduce the \qtraj code package allowing for the simulation of the quantum dynamics of heavy-quarkonium states in the quark-gluon plasma.  In practice, the code uses a stochastic unravelling of the Lindblad equation called the quantum trajectories algorithm to reduce the solution of the in-medium 3D Lindblad equation to the solution of a one-dimensional Schr\"odinger equation along with stochastically sampled quantum jumps which can change the color and angular momentum quantum numbers of the state.  The resulting 3D solution of the Lindblad equation allows one to compute the survival probabilities of heavy-quark bound states on a trajectory-by-trajectory basis.  For the most recent physics application of \qtraj\!\!, we refer the reader to~\cite{Brambilla:2021wkt}, where phenomenological predictions for bottomonium suppression and elliptic flow in 5 TeV Pb-Pb collisions are presented.

This paper provides detailed documentation for downloading, compiling, and running \qtraj in addition to providing theoretical background for the algorithm itself.  We compare \qtraj results obtained using a variety of lattice sizes, lattice spacings, temporal step sizes, etc., demonstrating that it is possible to achieve control of the various sources of systematic uncertainty.  We also compare run times for the \qtraj code to standard matrix-based solvers available in the QuTiP 2 package, finding that \qtraj allows one to efficiently solve the Lindblad equation on large lattices without having to impose an angular momentum cutoff.  The resulting code is embarrassingly parallel allowing the user to distribute the workload of generating independent quantum trajectories.  Associated with this paper, we publicly release the \qtraj code under a GPLv3 licence \cite{qtraj-download}.  In doing so, we seek to assist other researchers studying open quantum systems approaches to heavy-quarkonium dynamics in the QGP and other physically or theoretically similar systems.

\section*{Acknowledgments}

We thank N. Brambilla and A. Vairo for a very fruitful  collaboration.
M.A.E. received financial support from Xunta de Galicia (Centro singular de investigación de Galicia accreditation 2019-2022), the European Union ERDF, the “María de Maeztu” Units of Excellence program MDM2016-0692, the Spanish Research State Agency and from the
European Research Council project ERC-2018-ADG-835105 YoctoLHC.
M.S., A.I., and S.T. were supported by the U.S. Department of Energy, Office of Science, Office of Nuclear Physics Award No. DE-SC0013470. 
M.S., A.I., and S.T. also thank the Ohio Supercomputer Center for computational time under the auspices of Project No. PGS0253. 
J.H.W. was supported by the U.S. Department of Energy, Office of Science, Office of Nuclear Physics and Office of Advanced Scientific Computing Research within the framework of Scientific Discovery through Advance Computing (SciDAC) award Computing the Properties of Matter with Leadership Computing Resources. 
J.H.W.’s research was also funded by the Deutsche Forschungsgemeinschaft (DFG, German Research Foundation) - Projektnummer 417533893/GRK2575 ``Rethinking Quantum Field Theory''. 
P.V. acknowledges support from the Bundesministerium für Bildung und Forschung project no. 05P2018 and the DFG cluster of excellence ORIGINS funded by the Deutsche Forschungsgemeinschaft under Germany’s Excellence Strategy - EXC-2094-390783311. 

\appendix

\section{Packages required}
\label{sec:requirements}

In this appendix, we provide installation instructions for the \qtraj package.  The code has been tested on Mac OSX 10.15+, Ubuntu 18+, and Red Hat Enterprise Linux 7+.  Below we provide installation instructions for the former two since these are more commonly used.  To begin, we list the external packages required to compile and run {\tt QTraj}.  The first requirement is a modern C++ compiler, e.g. GNU g++ or LLVM.  In addition to this, \qtraj requires:
\begin{enumerate}
\item FFTW v3.3.8 or greater \cite{FFTW}.
\item GNU GSL v2.6 or greater \cite{GSL}.
\item Intel MKL v2021.2 or greater \cite{IntelOneAPI} {\bf OR} Armadillo 10.1.2 or greater \cite{Armadillo}.
\end{enumerate}
Compilation and running of the unit tests associated with \qtraj requires:
\begin{enumerate}
\item GoogleTest v1.10.0 or greater \cite{Googletest}.
\item Cmake v3.19 \cite{cmake} or greater is required for GoogleTest compilation.
\end{enumerate}

\subsection{Installation of FFTW}
\label{subsec:installfftw3}

\vspace{2mm}

For Mac OSX, we recommend the installation of the Homebrew package manager \cite{homebrew}.  Once this package manager is installed, the user should be able to execute the following commands in the terminal:
\begin{verbatim}
    $ brew install fftw3
\end{verbatim}
For Ubuntu, one can use {\tt apt-get}:
\begin{verbatim}
    $ sudo apt-get install fftw3-dev
\end{verbatim}

\subsection{Installation of GSL}
\label{subsec:installgsl}

\vspace{2mm}

For Mac OSX:
\begin{verbatim}
    $ brew install gsl
\end{verbatim}
For Ubuntu, one can use {\tt apt-get}:
\begin{verbatim}
    $ sudo apt-get install gsl-bin libgsl-dev
\end{verbatim}

\subsection{Installation of Intel MKL}
\label{subsec:installmkl}

\vspace{2mm}

To download Intel MKL, the user can follow the link in~\cite{IntelOneAPI} to download and run the full installer for their operating system. 
This requires registering for an account with Intel. Once the installer has completed installation, the following line must be added to the end of the user's {\tt .profile} or {\tt .bash\_profile} startup script; the terminal should be restarted afterwards.
\begin{verbatim}
    source /opt/intel/oneapi/setvars.sh
\end{verbatim}
If the MKL installation is in a non-standard location, the path to the {\tt compilervars.sh} script appearing above needs to be adjusted appropriately.

\subsection{Installation of Armadillo}
\label{subsec:installarmadillo}

\vspace{2mm}

For Mac OSX:
\begin{verbatim}
    $ brew install armadillo
\end{verbatim}
For Ubuntu, one can use {\tt apt-get}:
\begin{verbatim}
    $ sudo apt-get install liblapack-dev libblas-dev libboost-dev libarmadillo-dev
\end{verbatim}

\subsection{Installation of GoogleTest}
\label{subsec:installgtest}

\vspace{2mm}

For Mac OSX:
\begin{verbatim}
    $ brew install cmake
\end{verbatim}
For Ubuntu, one can use:
\begin{verbatim}
    $ sudo apt-get install cmake
\end{verbatim}
Once {\tt cmake} is installed, GoogleTest can be downloaded from~\cite{Googletest}, using

\begin{verbatim}
    $ git clone https://github.com/google/googletest
\end{verbatim}
or the user's preferred method. One must then execute:
\begin{verbatim}
    $ cd googletest
    $ mkdir build
    $ cd build
    $ cmake ..
    $ make
    $ sudo make install
\end{verbatim}

\section{Scripts for large scale deployment}
\label{sec:scripts}

The {\tt scripts} directory in the main package contains Torque Portable Batch System (PBS) \cite{openpbs} scripts and HTCondor \cite{condor} scripts that can assist with large scale deployment of the code in supercomputing environments:

\listingbox{%
\dirtree{%
.1 scripts.
.2 arrayJob.pbs.
.2 condorJob.sh.
.2 condor.job.
.2 scanImpactParameters.pbs.
.2 singleJob.pbs.
.2 manyTrajectories.pbs.
}
}

\section{Temperature file formats}
\label{sec:formats}

\qtraj v1.0 supports two temperature file formats.  To activate these options, the user can set the {\tt temperatureEvolution} parameter to 1 or 2.  For both options, the file to load is specified by the {\tt temperatureFile} parameter.

\subsection{Trajectory-averaged temperature format}
\label{sec:formats1}
\vspace{3mm}

For option 1, the file to which {\tt temperatureFile} points should have the following format
\begin{lstlisting}[language=bash]
# t [fm/c]  T [GeV]  alphax  alphay  alphaz  lambda [GeV]
<NUMBER OF TAU POINTS>
<TAU_1> <T_1> <ALPHAX_1> <ALPHAY_1> <ALPHAZ_1> <LAMBDA_1>
<TAU_2> <T_2> <ALPHAX_2> <ALPHAY_2> <ALPHAZ_2> <LAMBDA_2>
  .       .       .           .           .           . 
  .       .
  .       .       .           .           .           . 
\end{lstlisting}
where TAU is the proper time $\tau$, $T$ is the temperature, ALPHA is $\alpha_{x,y,z}$ and LAMBDA is $\lambda$, a temperature-like scale which corresponds to temperature in the equilibrium limit. 
The scale parameters $\alpha_{x,y,z}$ appear in the anisotropic distribution function \cite{Nopoush:2014pfa}
\begin{equation}
f(p) = f_{\rm iso} \left(\frac{1}{\lambda}\sqrt{\sum_i \frac{p^2_i}{\alpha^2_i}+m^2}\right)\,\, ,
\label{eq:faniso}
\end{equation}
where $i ~\epsilon~\{x,y,z\}$ and $f_{\rm iso}$ is an isotropic distribution function which in thermal equilibrium is given by a Bose-Einstein distribution.

\subsection{Trajectory-based temperature format}
\label{sec:formats2}
\vspace{3mm}

For option 2, the file to which {\tt temperatureFile} points should have the following format
\begin{lstlisting}[language=bash]
<NUMBER OF METADATA ELEMENTS>
<METADATA_1>
    .
    .
<METADATA_N>
<NUMBER OF TAU POINTS>
<TAU_1> <T_1>
<TAU_2> <T_2>
  .       .
  .       .
  .       .
\end{lstlisting}
As with option 1, $\tau$ is assumed to be in units of fm/c and $T$ is assumed to be in units of GeV.

In practice, the metadata values are floating point numbers that represent data about the given trajectory which can be used in post-processing.  Canonically, there are 6 metadata elements which record the following data for a given trajectory: 
\begin{enumerate}
    \item{Impact Parameter in fm.}
    \item{Initial production point $x$ in fm.}
    \item{Initial production point $y$ in fm.}
    \item{Initial point $\eta$ (spatial rapidity).}
    \item{Initial $p_T$ in GeV.}
    \item{Initial azimuthal angle $\phi$.}
\end{enumerate}

\section{Output file formats}
\label{sec:outputformats}

\subsection{Format of snapshot files}

If the user sets the {\tt saveWavefunctions} parameter to 1, snapshots of the wave-function are saved periodically to disk in files named {\tt snapshot\_<number>.tsv} where the number is the time step of the snapshot.  The format of these files is as follows
\begin{lstlisting}[language=bash]
<R_1> <ABS_PSI_1> <RE_PSI_1> <IM_PSI_1>
<R_2> <ABS_PSI_2> <RE_PSI_2> <IM_PSI_2>
  .       
  .
  .
\end{lstlisting}
where {\tt R} is the position in 1/GeV, {\tt ABS\_PSI} is $|\psi|$, {\tt RE\_PSI} is $\Re[\psi]$, and {\tt IM\_PSI} is $\Im[\psi]$.

\subsection{Format of summary.tsv}

If the user sets the {\tt outputSummaryFile} variable to 1, \qtraj outputs summary information concerning the evolution.  Typically, this output is only used for debugging on a single quantum trajectory basis.  The columns in this file correspond to
\begin{lstlisting}[language=bash]
<TIME_1> <NORM_1> <COLOR_1> <L_1> <1S_1> <2S_1> <1P_1> <3S_1> <2P_1> <1D_1> <R_1> 
<TIME_2> <NORM_2> <COLOR_2> <L_2> <1S_2> <2S_2> <1P_2> <3S_2> <2P_2> <1D_2> <R_2> 
  .       
  .
  .
\end{lstlisting}
where the following notation has been used: 
\begin{enumerate}
\item
\texttt{TIME} is the simulation time $t$ in fm.
\item
\texttt{NORM} is the norm of the wave-function $\braket{\psi(t)|\psi(t)}$.
\item
\texttt{COLOR} is the color state; 0 = singlet, 1 = octet.
\item
\texttt{L} is the angular momentum quantum number.
\item
\texttt{1S}, \texttt{2S}, \texttt{1P} etc. are the quantum mechanical overlaps, e.g. $|\braket{1S|\psi(t)}|^2$.
\item
\texttt{R} is the expectation value of $r$ normalized by the 1S expectation value of $r$, $\bra{\psi(t)} r \ket{\psi(t)}/\bra{1S} r \ket{1S}$.
\end{enumerate}

\subsection{Format of ratios.tsv}

If the user sets the {\tt temperatureEvolution} parameter to 1 or 2, the output file {\tt output/ratios.tsv} contains rows of data in the following format 
\begin{lstlisting}[language=bash]
<RAT_1S_1> <RAT_2S_1> <RAT_1P_1> <RAT_3S_1> <RAT_2P_1> <RAT_1D_1> <RAND_1> <INIT_L_1> 
<RAT_1S_2> <RAT_2S_2> <RAT_1P_2> <RAT_3S_2> <RAT_2P_2> <RAT_1D_2> <RAND_2> <INIT_L_2>  
  .       
  .
  .
\end{lstlisting}

If the user sets the temperatureEvolution parameter to 3, then the output file {\tt output/ratios.tsv} will contain rows of data in the following format 
\begin{lstlisting}[language=bash]
<IP_1> <X_1>  <Y_1> <ETA_1> <PT_1> <PHI_1>
<RAT_1S_1> <RAT_2S_1> <RAT_1P_1> <RAT_3S_1> <RAT_2P_1> <RAT_1D_1> <RAND_1> <INIT_L_1> 
<IP_2> <X_2>  <Y_2> <ETA_2> <PT_2> <PHI_2>
<RAT_1S_2> <RAT_2S_2> <RAT_1P_2> <RAT_3S_2> <RAT_2P_2> <RAT_1D_2> <RAND_2> <INIT_L_2>  
  .       
  .
  .
\end{lstlisting}
where the following abbreviations have been used: 
\begin{enumerate}
\item
\texttt{RAT} is the overlap ratio or raw overlap depending on the state in consideration.\footnote{Which output form is given for a given state depends on the initial overlap of the state.  In the case that there is a zero overlap in the initial condition, \qtraj records the final quantum mechanical overlap modulus squared, otherwise the ratio of the final to initial quantum mechanical overlaps is recorded.}
\item
\texttt{RAND} is the first random number generated during the evolution.
\item
\texttt{INIT\_L} is the initial value of angular momentum quantum number $l$.
\item
\texttt{IP} is the impact parameter in fm.
\item
\texttt{X}, \texttt{Y} are the initial production point $(X,Y)$ (in fm) for the physical trajectory.
\item
\texttt{ETA} is the spatial rapidity $\eta$.
\item
\texttt{PT} is the initial transverse momentum $p_T$ in GeV for the physical trajectory.
\item
\texttt{PHI} is the initial azimuthal angle $\phi$ of the physical trajectory.
\end{enumerate}

\section{Supported potentials}
\label{sec:potential}

\subsection{Munich potential}
\label{sec:munichpotential}
From Eq. (\ref{eq:hamiltonian}), we can write down the combined Hamiltonian \cite{Brambilla:2017zei} for singlet and octet states as 
\begin{equation}
  H = \left(\begin{array}{c c}
h_s + \frac{1}{2}\,\gamma\,r^2& 0\\
0 & h_o + \frac{7}{32}\,\gamma\,r^2
\end{array}\right)\,\,,\label{E.1}
\end{equation}
where
\begin{equation}
    h_s = \frac{\mathbf{p}^2}{m} - C_F \frac{\alpha_s}{r} = \frac{\mathbf{p}^2}{m} - \frac{\alpha}{r}\,\,,
\end{equation}
\begin{equation}
  h_o = \frac{\mathbf{p}^2}{m} + \frac{\alpha_s}{2 N_c r}=\frac{\mathbf{p}^2}{m} + \frac{1}{8} \frac{\alpha}{r}\,\,.
\end{equation}
with,  $N_c = 3$, $C_F = \frac{N^2_c - 1}{2N_c} = \frac{4}{3}$ , and $\alpha = C_F \,\alpha_s = \frac{4}{3}\alpha_s$.

From Eq. (\ref{E.1}), we can write down the singlet Hamiltonian as 
\begin{equation}
  H_s = h_s + \frac{1}{2}\,\gamma\,r^2 =\frac{\mathbf{p}^2}{m} -  \frac{\alpha}{r} + \frac{1}{2}\,\hat{\gamma}\,T^3 r^2 \,\,, \label{E.4}
\end{equation}
and the octet Hamiltonian as 
\begin{equation}
  H_o = h_o + \frac{7}{32}\,\gamma\,r^2 = \frac{\mathbf{p}^2}{m} +  \frac{1}{8}\frac{\alpha}{r} + \frac{7}{32}\,\hat{\gamma}\,T^3 r^2 \,\,,\label{E.5}
\end{equation}
where $\gamma = \hat{\gamma}T^3$ with $-3.5 < \hat{\gamma} < 0$ \cite{Brambilla:2019tpt,Brambilla:2020qwo}. The factor $\frac{7}{32}$ arises as $\frac{1}{4} \frac{N^2_c-2}{N^2_c-1}$.

\begin{figure}[t]
\begin{center}
\includegraphics[width=0.45\linewidth]{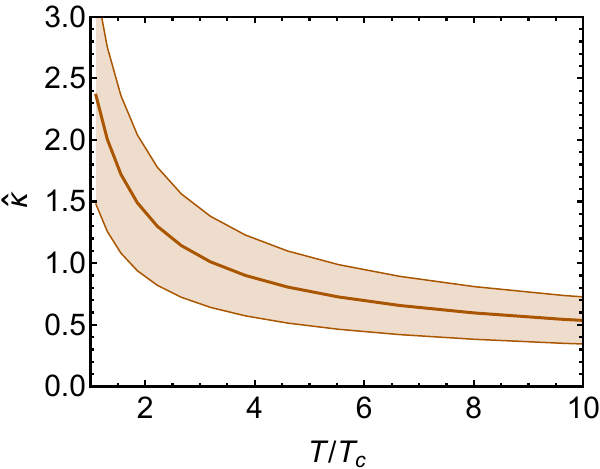}
\end{center}
\caption{$\hat\kappa$ as a function of the temperature in units of $T_c$, where $T_c$ = 155 MeV; based on data from~\cite{Brambilla:2020siz}.}
\label{fig:kappaHat}
\end{figure}

From Eq. (\ref{E.4}) and Eq. (\ref{E.5}), we extract the real parts of the singlet and octet potentials %
\begin{equation}
V_R^s  (r) = -  \frac{\alpha}{r} + \frac{1}{2}\,\hat{\gamma}\,T^3 r^2 \,\,,
\end{equation}
\begin{equation}
V_R^o  (r) = \frac{1}{8}\frac{\alpha}{r} + \frac{7}{32}\,\hat{\gamma}\,T^3 r^2 \,\,.
\end{equation}
We use Eq. (C3) and Eq. (C6) of~\cite{Brambilla:2017zei}  to write down the imaginary part of the singlet and octet potentials 
\begin{equation}
V_I^s  (r) = -\frac{1}{2}~\kappa ~r^2 = -\frac{1}{2}~\hat{\kappa}~T^3 r^2\,\,,
\end{equation}
and
\begin{equation}
V_I^o (r)  = -\frac{1}{4} \frac{N^2_c-2}{N^2_c-1}~\kappa ~r^2 = -\frac{7}{32}~\kappa ~r^2 = -\frac{7}{32}~\hat{\kappa}~T^3 r^2\,\,,
\end{equation}
where $\kappa = \hat{\kappa} T^3$.  The next-to-leading-order perturbative expansion of $\hat{\kappa}$ is defined by Eq. (C4) of~\cite{Brambilla:2017zei}. We use a parameterization of this result which is fitted to lattice data~\cite{Brambilla:2020siz}. The resulting parameterization for $\hat{\kappa}$ is given by \cite{Brambilla:2017zei}
\be
\hat{\kappa} =
\begin{cases}  
	\left(0.00397x^{3/2} - 0.08341x + 0.98031x^{1/2} - 0.51889\right)^{-1} &\mbox{in case of central limit fit} \\
	\left(0.00712x^{3/2} - 0.14262x + 1.55601x^{1/2} - 0.80712\right)^{-1} &\mbox{in case of lower limit fit} \\
	\left(0.00272x^{3/2} - 0.05854x + 0.71513x^{1/2} - 0.38197\right)^{-1} &\mbox{in case of upper limit fit} 
\end{cases} \, ,
\ee
where $x = T/T_C$. In Fig.~\ref{fig:kappaHat}, we plot $\hat\kappa$ as a function of the temperature in units of $T_c$ where $T_c$ = 155 MeV, which is close to recent determinations in (2+1)-flavor lattice QCD with physical quark masses, 
e.g.~\cite{Bazavov:2016uvm, Bazavov:2018mes, Borsanyi:2020fev}.

The resulting complex-valued potentials for singlet and octet states are 
\be
V^s_{\rm Munich} (r) = V^s_R (r) + i V^s_I (r) \, , 
\ee
and
\be
V^o_{\rm Munich} (r) = V^o_R (r) + i V^o_I (r) \, .
\ee

\subsection{Isotropic KSU potential}
\label{sec:isoksupotential}

For the KSU potential, we make use of a Karsch-Mehr-Satz type potential based on the internal energy~\cite{Karsch:1987pv,Dumitru:2009ni,Strickland:2011aa}
\begin{equation}
V^{\mathrm{iso}}_{\mathrm{KMS}}(r)=-\frac{a}{r}(1+ m_D r)e^{-m_D  r}+\frac{2\sigma}{m_D }\left[1-e^{-m_D  r}\right]-\sigma r e^{-m_D  r} ~,
\end{equation}
where \mbox{$a = 0.409$} is the effective coupling, \mbox{$\sigma = 0.21$~GeV$^2$} is the string tension, and the isotropic Debye mass is defined by 
\begin{equation}
m_D(T)=F_{m_D}T\sqrt{N_c\left(1+\frac{N_F}{6}\right)\pi\alpha}~,~~\alpha = \frac{4}{3}\alpha_s (2\pi T)~,
\end{equation}
where $F_{m_D}$ is an adjustable factor called the $m_D$-factor and $\alpha_s(5\text{ GeV}) = 0.2034$~\cite{McNeile:2010ji} is the strong coupling.  Note that, in the current version of \qtraj\!, the various parameters such as $\sigma$, etc. are hardcoded as defined in {\tt potential.h}.  Changing these parameters currently requires recompilation of the code.

For the imaginary part of the model potential, we use the result obtained from leading-order finite temperature perturbation theory~\cite{Laine:2006ns} assuming the hierarchy $r \sim 1/m_D$ of the chromoelectric screening regime,
\begin{equation} 
V^{\mathrm{iso}}_{\mathrm{I}}(r) = - \alpha_s C_F T \phi(\hat{r}) \, ,
\end{equation}
where $\hat{r} \equiv m_D r$ and
\begin{equation}
 \phi(\hat{r}) \equiv 2\int_0^{\infty}dz \frac{
z}{(z^2+1)^2} \left[1-\frac{\sin(z\, \hat{r})}{z\, \hat{r}}\right] .
\end{equation}
In vacuum, we take the heavy-quark potential to be given by a Cornell potential
\be
V_{\rm vac}(r) =
\begin{cases}  
	-\frac{a}{r} + \sigma r &\mbox{if } r \leq r_{\rm SB} \\
	-\frac{a}{r_{\rm SB} } + \sigma r_{\rm SB}   & \mbox{if } r > r_{\rm SB}
\end{cases} \, ,
\label{eq:vvac}
\ee
where \mbox{$r_{\rm SB}  =$ 1.25 fm} is the finite string breaking distance.
To match smoothly onto the zero temperature limit, we use
\begin{equation}
V^{\rm iso}_R (r)=
\begin{cases}  V^{\rm iso}_{\rm KMS}(r)  &\mbox{if } V^{\rm iso}_{\rm KMS}(r)  \leq V_{\rm vac}(r_{\rm SB}) \\
V_{\rm vac}(r_{\rm SB}) & \mbox{if } V^{\rm iso}_{\rm KMS}(r) > V_{\rm vac}(r_{\rm SB})
\end{cases} \, .
\end{equation}

The resulting final isotropic complex-valued KSU potential is of the form 
\begin{equation}
V^{\rm iso}_{\rm KSU} (r) = V^{\rm iso}_R (r) + i V^{\rm iso}_I (r) \, .
\end{equation}

\subsection{Anisotropic KSU potential}
\label{sec:anisoksupotential}

After considering the ``minimal'' extension \cite{Dumitru:2009ni} of the internal energy based isotropic KMS potential \cite{Karsch:1987pv,Strickland:2011aa}, we write down the real part of the anisotropic KMS potential
\begin{equation}
V^{\mathrm{aniso}}_{\mathrm{KMS}}(r)=-\frac{a}{r}(1+\mu r)e^{-\mu r}+\frac{2\sigma}{\mu}\left[1-e^{-\mu r}\right]-\sigma r e^{-\mu r} ~,
\end{equation}
where $\mu$ is the anisotropic screening scale (anisotropic Debye mass) defined via
\begin{equation}
\mu^2(\xi, T) = f(\xi) m^2_D(T) ~.\label{mumd}
\end{equation}
The function $f(\xi)$ in Eq.(\ref{mumd}) is given by \cite{Strickland:2018ayk} 
\begin{equation}
f(\xi)=\frac{\sqrt{2}(1+\xi)\arcsin{\left(\sqrt{\frac{\xi}{1+\xi}}\right)}}{\sqrt{\xi}\sqrt{1+\xi+\frac{(1+\xi)^2 \arcsin{\left(\sqrt{\frac{\xi}{1+\xi}}\right)}}{\sqrt{\xi}}}}~,\label{ffunc}
\end{equation}
which has the following properties
\begin{equation}
\lim_{\xi \rightarrow 0}f(\xi) = 1-\frac{2}{45}\xi^2 + \frac{44}{945}\xi^3 - \frac{8}{189}\xi^4 + {\cal O}\left(\xi\right)^{9/2}~,
\end{equation}
and
\begin{equation}
\lim_{\xi \rightarrow \infty}f(\xi) = \sqrt{\pi}\left(\frac{1}{\xi}\right)^{1/4} - \frac{2}{\sqrt{\pi}}\left(\frac{1}{\xi}\right)^{3/4} + \frac{4}{3\sqrt{\pi}}\left(\frac{1}{\xi}\right)^{7/4} + {\cal O}\left(\frac{1}{\xi}\right)^{9/4}~,
\end{equation}
in the small and large-$\xi$ limit respectively.

When $\xi$ is negative, we use 
\begin{equation}
f(\xi) = \frac{\sqrt{2}\arcsinh{\left(\sqrt{\frac{\abs{\xi}}{1-\abs{\xi}}}~\right)}}{\abs{\xi}^{1/4}\sqrt{\frac{\sqrt{\abs{\xi}}}{1-\abs{\xi}}+\arcsinh{\left(\sqrt{\frac{\abs{\xi}}{1-\abs{\xi}}}~\right)}}}~~,~\mathrm{when}~~\xi< 0 ~.
\end{equation}
For the imaginary part of the potential, we use \cite{islamforth}
\begin{equation}
V^{\rm aniso}_I (r) = - \frac{\alpha_s C_F \lambda \phi (\hat{r})}{ \sqrt{1 + \frac{\xi}{3}} } \, ,
\label{eq:allanisoimv}
\end{equation}
with $\hat{r} = m_D(\lambda) r$ and $m_D(\lambda)=F_{m_D}\lambda~\sqrt{N_c\left(1+\frac{N_F}{6}\right)\pi\alpha}$ .
The parameter $\lambda$ is a temperature-like scale, see \ref{sec:formats1} and, in particular, Eq.~\ref{eq:faniso} for its functional definition which is based on a generalization of the momentum-anisotropic form introduced originally in~\cite{Romatschke:2003ms} and extended to include multiple anisotropy parameters in~\cite{Martinez:2012tu,Nopoush:2014pfa}.

To match smoothly onto the zero temperature limit, we use
\begin{equation}
V^{\rm aniso}_R (r)=
\begin{cases}  V^{\rm aniso}_{\rm KMS}(r)  &\mbox{if } V^{\rm aniso}_{\rm KMS}(r)  \leq V_{\rm vac}(r_{\rm SB}) \\
V_{\rm vac}(r_{\rm SB}) & \mbox{if } V^{\rm aniso}_{\rm KMS}(r) > V_{\rm vac}(r_{\rm SB})
\end{cases} \, .
\label{eq:vmedre}
\end{equation}
The resulting final anisotropic complex-valued KSU potential is of the form 
\begin{equation}
V^{\rm aniso}_{\rm KSU}  (r) = V^{\rm aniso}_R (r) + i V^{\rm aniso}_I (r) \, .
\label{eq:vform}
\end{equation}
%

\bibliographystyle{model1-num-names}
\bibliography{qtraj-fftw}

\begin{thebibliography}{87}
\expandafter\ifx\csname natexlab\endcsname\relax\def\natexlab#1{#1}\fi
\providecommand{\url}[1]{\texttt{#1}}
\providecommand{\href}[2]{#2}
\providecommand{\path}[1]{#1}
\providecommand{\DOIprefix}{doi:}
\providecommand{\ArXivprefix}{arXiv:}
\providecommand{\URLprefix}{URL: }
\providecommand{\Pubmedprefix}{pmid:}
\providecommand{\doi}[1]{\href{http://dx.doi.org/#1}{\path{#1}}}
\providecommand{\Pubmed}[1]{\href{pmid:#1}{\path{#1}}}
\providecommand{\bibinfo}[2]{#2}
\ifx\xfnm\relax \def\xfnm[#1]{\unskip,\space#1}\fi
\bibitem[{Bazavov(2013)}]{Bazavov:2013txa}
\bibinfo{author}{A.~Bazavov},
\newblock \bibinfo{title}{{An overview of (selected) recent results in
  finite-temperature lattice QCD}},
\newblock \bibinfo{journal}{J. Phys. Conf. Ser.} \bibinfo{volume}{446}
  (\bibinfo{year}{2013}) \bibinfo{pages}{012011}.
\bibitem[{Borsanyi(2017)}]{Borsanyi:2016bzg}
\bibinfo{author}{S.~Borsanyi},
\newblock \bibinfo{title}{{Frontiers of finite temperature lattice QCD}},
\newblock \bibinfo{journal}{EPJ Web Conf.} \bibinfo{volume}{137}
  (\bibinfo{year}{2017}) \bibinfo{pages}{01006}.
\bibitem[{Bazavov and Weber(2021)}]{Bazavov:2020teh}
\bibinfo{author}{A.~Bazavov}, \bibinfo{author}{J.~H. Weber},
\newblock \bibinfo{title}{{Color Screening in Quantum Chromodynamics}},
\newblock \bibinfo{journal}{Prog. Part. Nucl. Phys.} \bibinfo{volume}{116}
  (\bibinfo{year}{2021}) \bibinfo{pages}{103823}.
\bibitem[{Averbeck et~al.(2015)Averbeck, Harris, and Schenke}]{Averbeck2015}
\bibinfo{author}{R.~Averbeck}, \bibinfo{author}{J.~W. Harris},
  \bibinfo{author}{B.~Schenke}, \bibinfo{title}{{Heavy-Ion Physics at the
  LHC}}, \bibinfo{publisher}{Springer International Publishing},
  \bibinfo{address}{Cham}, \bibinfo{year}{2015}, pp. \bibinfo{pages}{355--420}.
\bibitem[{Busza et~al.(2018)Busza, Rajagopal, and van~der Schee}]{Busza:2018}
\bibinfo{author}{W.~Busza}, \bibinfo{author}{K.~Rajagopal},
  \bibinfo{author}{W.~van~der Schee},
\newblock \bibinfo{title}{Heavy ion collisions: The big picture and the big
  questions},
\newblock \bibinfo{journal}{Annual Review of Nuclear and Particle Science}
  \bibinfo{volume}{68} (\bibinfo{year}{2018}) \bibinfo{pages}{339--376}.
\bibitem[{Matsui and Satz(1986)}]{Matsui:1986dk}
\bibinfo{author}{T.~Matsui}, \bibinfo{author}{H.~Satz},
\newblock \bibinfo{title}{{$J/\psi$ Suppression by Quark-Gluon Plasma
  Formation}},
\newblock \bibinfo{journal}{Phys. Lett.} \bibinfo{volume}{B178}
  (\bibinfo{year}{1986}) \bibinfo{pages}{416}.
\bibitem[{Karsch et~al.(1988)Karsch, Mehr, and Satz}]{Karsch:1987pv}
\bibinfo{author}{F.~Karsch}, \bibinfo{author}{M.~T. Mehr},
  \bibinfo{author}{H.~Satz},
\newblock \bibinfo{title}{{Color Screening and Deconfinement for Bound States
  of Heavy Quarks}},
\newblock \bibinfo{journal}{Z. Phys.} \bibinfo{volume}{C37}
  (\bibinfo{year}{1988}) \bibinfo{pages}{617}.
\bibitem[{Laine et~al.(2007)Laine, Philipsen, Romatschke, and
  Tassler}]{Laine:2006ns}
\bibinfo{author}{M.~Laine}, \bibinfo{author}{O.~Philipsen},
  \bibinfo{author}{P.~Romatschke}, \bibinfo{author}{M.~Tassler},
\newblock \bibinfo{title}{{Real-time static potential in hot QCD}},
\newblock \bibinfo{journal}{JHEP} \bibinfo{volume}{03} (\bibinfo{year}{2007})
  \bibinfo{pages}{054}.
\bibitem[{Strickland(2011)}]{Strickland:2011mw}
\bibinfo{author}{M.~Strickland},
\newblock \bibinfo{title}{{Thermal Upsilon(1s) and chi-b1 suppression in
  sqrt(s-NN)=2.76 TeV Pb-Pb collisions at the LHC}},
\newblock \bibinfo{journal}{Phys.Rev.Lett.} \bibinfo{volume}{107}
  (\bibinfo{year}{2011}) \bibinfo{pages}{132301}.
\bibitem[{Rothkopf(2020)}]{Rothkopf:2019ipj}
\bibinfo{author}{A.~Rothkopf},
\newblock \bibinfo{title}{{Heavy Quarkonium in Extreme Conditions}},
\newblock \bibinfo{journal}{Phys. Rept.} \bibinfo{volume}{858}
  (\bibinfo{year}{2020}) \bibinfo{pages}{1--117}.
\bibitem[{Akamatsu(2020)}]{Akamatsu:2020ypb}
\bibinfo{author}{Y.~Akamatsu},
\newblock \bibinfo{title}{{Quarkonium in Quark-Gluon Plasma: Open Quantum
  System Approaches Re-examined}}  (\bibinfo{year}{2020}).
\bibitem[{Rothkopf(2021)}]{Rothkopf:2020vfz}
\bibinfo{author}{A.~Rothkopf},
\newblock \bibinfo{title}{{Quarkonium production and suppression: Theory}},
\newblock \bibinfo{journal}{Nucl. Phys. A} \bibinfo{volume}{1005}
  (\bibinfo{year}{2021}) \bibinfo{pages}{121922}.
\bibitem[{Brambilla et~al.(2008)Brambilla, Ghiglieri, Vairo, and
  Petreczky}]{Brambilla:2008cx}
\bibinfo{author}{N.~Brambilla}, \bibinfo{author}{J.~Ghiglieri},
  \bibinfo{author}{A.~Vairo}, \bibinfo{author}{P.~Petreczky},
\newblock \bibinfo{title}{Static quark-antiquark pairs at finite temperature},
\newblock \bibinfo{journal}{Phys. Rev.} \bibinfo{volume}{D78}
  (\bibinfo{year}{2008}) \bibinfo{pages}{014017}.
\bibitem[{Brambilla et~al.(2011)Brambilla, Escobedo, Ghiglieri, and
  Vairo}]{Brambilla:2011sg}
\bibinfo{author}{N.~Brambilla}, \bibinfo{author}{M.~A. Escobedo},
  \bibinfo{author}{J.~Ghiglieri}, \bibinfo{author}{A.~Vairo},
\newblock \bibinfo{title}{{Thermal width and gluo-dissociation of quarkonium in
  pNRQCD}},
\newblock \bibinfo{journal}{JHEP} \bibinfo{volume}{12} (\bibinfo{year}{2011})
  \bibinfo{pages}{116}.
\bibitem[{Brambilla et~al.(2013)Brambilla, Escobedo, Ghiglieri, and
  Vairo}]{Brambilla:2013dpa}
\bibinfo{author}{N.~Brambilla}, \bibinfo{author}{M.~A. Escobedo},
  \bibinfo{author}{J.~Ghiglieri}, \bibinfo{author}{A.~Vairo},
\newblock \bibinfo{title}{{Thermal width and quarkonium dissociation by
  inelastic parton scattering}},
\newblock \bibinfo{journal}{JHEP} \bibinfo{volume}{05} (\bibinfo{year}{2013})
  \bibinfo{pages}{130}.
\bibitem[{Braun-Munzinger and Stachel(2000)}]{BraunMunzinger:2000px}
\bibinfo{author}{P.~Braun-Munzinger}, \bibinfo{author}{J.~Stachel},
\newblock \bibinfo{title}{{(Non)thermal aspects of charmonium production and a
  new look at J / psi suppression}},
\newblock \bibinfo{journal}{Phys. Lett. B} \bibinfo{volume}{490}
  (\bibinfo{year}{2000}) \bibinfo{pages}{196--202}.
\bibitem[{Thews et~al.(2001)Thews, Schroedter, and Rafelski}]{Thews:2000rj}
\bibinfo{author}{R.~L. Thews}, \bibinfo{author}{M.~Schroedter},
  \bibinfo{author}{J.~Rafelski},
\newblock \bibinfo{title}{{Enhanced $J/\psi$ production in deconfined quark
  matter}},
\newblock \bibinfo{journal}{Phys. Rev.} \bibinfo{volume}{C63}
  (\bibinfo{year}{2001}) \bibinfo{pages}{054905}.
\bibitem[{Emerick et~al.(2012)Emerick, Zhao, and Rapp}]{Emerick:2011xu}
\bibinfo{author}{A.~Emerick}, \bibinfo{author}{X.~Zhao},
  \bibinfo{author}{R.~Rapp},
\newblock \bibinfo{title}{{Bottomonia in the Quark-Gluon Plasma and their
  Production at RHIC and LHC}},
\newblock \bibinfo{journal}{Eur. Phys. J.} \bibinfo{volume}{A48}
  (\bibinfo{year}{2012}) \bibinfo{pages}{72}.
\bibitem[{Borghini and Gombeaud(2012)}]{Borghini:2011ms}
\bibinfo{author}{N.~Borghini}, \bibinfo{author}{C.~Gombeaud},
\newblock \bibinfo{title}{{Heavy quarkonia in a medium as a quantum dissipative
  system: Master equation approach}},
\newblock \bibinfo{journal}{Eur. Phys. J. C} \bibinfo{volume}{72}
  (\bibinfo{year}{2012}) \bibinfo{pages}{2000}.
\bibitem[{Akamatsu(2013)}]{Akamatsu:2012vt}
\bibinfo{author}{Y.~Akamatsu},
\newblock \bibinfo{title}{{Real-time quantum dynamics of heavy quark systems at
  high temperature}},
\newblock \bibinfo{journal}{Phys. Rev.} \bibinfo{volume}{D87}
  (\bibinfo{year}{2013}) \bibinfo{pages}{045016}.
\bibitem[{Akamatsu(2015)}]{Akamatsu:2014qsa}
\bibinfo{author}{Y.~Akamatsu},
\newblock \bibinfo{title}{{Heavy quark master equations in the Lindblad form at
  high temperatures}},
\newblock \bibinfo{journal}{Phys. Rev.} \bibinfo{volume}{D91}
  (\bibinfo{year}{2015}) \bibinfo{pages}{056002}.
\bibitem[{Blaizot et~al.(2016)Blaizot, De~Boni, Faccioli, and
  Garberoglio}]{Blaizot:2015hya}
\bibinfo{author}{J.-P. Blaizot}, \bibinfo{author}{D.~De~Boni},
  \bibinfo{author}{P.~Faccioli}, \bibinfo{author}{G.~Garberoglio},
\newblock \bibinfo{title}{{Heavy quark bound states in a
  quark\textendash{}gluon plasma: Dissociation and recombination}},
\newblock \bibinfo{journal}{Nucl. Phys. A} \bibinfo{volume}{946}
  (\bibinfo{year}{2016}) \bibinfo{pages}{49--88}.
\bibitem[{Katz and Gossiaux(2016)}]{Katz:2015qja}
\bibinfo{author}{R.~Katz}, \bibinfo{author}{P.~B. Gossiaux},
\newblock \bibinfo{title}{{The Schrödinger–Langevin equation with and
  without thermal fluctuations}},
\newblock \bibinfo{journal}{Annals Phys.} \bibinfo{volume}{368}
  (\bibinfo{year}{2016}) \bibinfo{pages}{267--295}.
\bibitem[{Brambilla et~al.(2017)Brambilla, Escobedo, Soto, and
  Vairo}]{Brambilla:2016wgg}
\bibinfo{author}{N.~Brambilla}, \bibinfo{author}{M.~A. Escobedo},
  \bibinfo{author}{J.~Soto}, \bibinfo{author}{A.~Vairo},
\newblock \bibinfo{title}{{Quarkonium suppression in heavy-ion collisions: an
  open quantum system approach}},
\newblock \bibinfo{journal}{Phys. Rev.} \bibinfo{volume}{D96}
  (\bibinfo{year}{2017}) \bibinfo{pages}{034021}.
\bibitem[{Brambilla et~al.(2018)Brambilla, Escobedo, Soto, and
  Vairo}]{Brambilla:2017zei}
\bibinfo{author}{N.~Brambilla}, \bibinfo{author}{M.~A. Escobedo},
  \bibinfo{author}{J.~Soto}, \bibinfo{author}{A.~Vairo},
\newblock \bibinfo{title}{{Heavy quarkonium suppression in a fireball}},
\newblock \bibinfo{journal}{Phys. Rev.} \bibinfo{volume}{D97}
  (\bibinfo{year}{2018}) \bibinfo{pages}{074009}.
\bibitem[{Kajimoto et~al.(2018)Kajimoto, Akamatsu, Asakawa, and
  Rothkopf}]{Kajimoto:2017rel}
\bibinfo{author}{S.~Kajimoto}, \bibinfo{author}{Y.~Akamatsu},
  \bibinfo{author}{M.~Asakawa}, \bibinfo{author}{A.~Rothkopf},
\newblock \bibinfo{title}{{Dynamical dissociation of quarkonia by wave function
  decoherence}},
\newblock \bibinfo{journal}{Phys. Rev.} \bibinfo{volume}{D97}
  (\bibinfo{year}{2018}) \bibinfo{pages}{014003}.
\bibitem[{Blaizot and Escobedo(2018)}]{Blaizot:2017ypk}
\bibinfo{author}{J.-P. Blaizot}, \bibinfo{author}{M.~A. Escobedo},
\newblock \bibinfo{title}{{Quantum and classical dynamics of heavy quarks in a
  quark-gluon plasma}},
\newblock \bibinfo{journal}{JHEP} \bibinfo{volume}{06} (\bibinfo{year}{2018})
  \bibinfo{pages}{034}.
\bibitem[{Akamatsu et~al.(2018)Akamatsu, Asakawa, Kajimoto, and
  Rothkopf}]{Akamatsu:2018xim}
\bibinfo{author}{Y.~Akamatsu}, \bibinfo{author}{M.~Asakawa},
  \bibinfo{author}{S.~Kajimoto}, \bibinfo{author}{A.~Rothkopf},
\newblock \bibinfo{title}{{Quantum dissipation of a heavy quark from a
  nonlinear stochastic Schr\"odinger equation}},
\newblock \bibinfo{journal}{JHEP} \bibinfo{volume}{07} (\bibinfo{year}{2018})
  \bibinfo{pages}{029}.
\bibitem[{Yao and Mehen(2019)}]{Yao:2018nmy}
\bibinfo{author}{X.~Yao}, \bibinfo{author}{T.~Mehen},
\newblock \bibinfo{title}{{Quarkonium in-medium transport equation derived from
  first principles}},
\newblock \bibinfo{journal}{Phys. Rev. D} \bibinfo{volume}{99}
  (\bibinfo{year}{2019}) \bibinfo{pages}{096028}.
\bibitem[{Yao and M\"uller(2019)}]{Yao:2018sgn}
\bibinfo{author}{X.~Yao}, \bibinfo{author}{B.~M\"uller},
\newblock \bibinfo{title}{{Quarkonium inside the quark-gluon plasma: Diffusion,
  dissociation, recombination, and energy loss}},
\newblock \bibinfo{journal}{Phys. Rev. D} \bibinfo{volume}{100}
  (\bibinfo{year}{2019}) \bibinfo{pages}{014008}.
\bibitem[{Blaizot and Escobedo(2018)}]{Blaizot:2018oev}
\bibinfo{author}{J.-P. Blaizot}, \bibinfo{author}{M.~A. Escobedo},
\newblock \bibinfo{title}{{Approach to equilibrium of a quarkonium in a
  quark-gluon plasma}},
\newblock \bibinfo{journal}{Phys. Rev.} \bibinfo{volume}{D98}
  (\bibinfo{year}{2018}) \bibinfo{pages}{074007}.
\bibitem[{Yao and Mehen(2020)}]{Yao:2020eqy}
\bibinfo{author}{X.~Yao}, \bibinfo{author}{T.~Mehen},
\newblock \bibinfo{title}{{Quarkonium Semiclassical Transport in Quark-Gluon
  Plasma: Factorization and Quantum Correction}},
\newblock \bibinfo{journal}{JHEP} \bibinfo{volume}{21} (\bibinfo{year}{2020})
  \bibinfo{pages}{062}.
\bibitem[{Yao(2021)}]{Yao:2021lus}
\bibinfo{author}{X.~Yao},
\newblock \bibinfo{title}{{Open Quantum Systems for Quarkonia}}
  (\bibinfo{year}{2021}).
\bibitem[{Gorini et~al.(1976)Gorini, Kossakowski, and
  Sudarshan}]{Gorini:1975nb}
\bibinfo{author}{V.~Gorini}, \bibinfo{author}{A.~Kossakowski},
  \bibinfo{author}{E.~Sudarshan},
\newblock \bibinfo{title}{{Completely Positive Dynamical Semigroups of N Level
  Systems}},
\newblock \bibinfo{journal}{J. Math. Phys.} \bibinfo{volume}{17}
  (\bibinfo{year}{1976}) \bibinfo{pages}{821}.
\bibitem[{Lindblad(1976)}]{Lindblad:1975ef}
\bibinfo{author}{G.~Lindblad},
\newblock \bibinfo{title}{{On the Generators of Quantum Dynamical Semigroups}},
\newblock \bibinfo{journal}{Commun. Math. Phys.} \bibinfo{volume}{48}
  (\bibinfo{year}{1976}) \bibinfo{pages}{119}.
\bibitem[{Johansson et~al.(2012)Johansson, Nation, and Nori}]{qutip1}
\bibinfo{author}{J.~Johansson}, \bibinfo{author}{P.~Nation},
  \bibinfo{author}{F.~Nori},
\newblock \bibinfo{title}{Qutip: An open-source python framework for the
  dynamics of open quantum systems},
\newblock \bibinfo{journal}{Computer Physics Communications}
  \bibinfo{volume}{183} (\bibinfo{year}{2012}) \bibinfo{pages}{1760--1772}.
\bibitem[{Johansson et~al.(2013)Johansson, Nation, and Nori}]{qutip2}
\bibinfo{author}{J.~Johansson}, \bibinfo{author}{P.~Nation},
  \bibinfo{author}{F.~Nori},
\newblock \bibinfo{title}{Qutip 2: A python framework for the dynamics of open
  quantum systems},
\newblock \bibinfo{journal}{Computer Physics Communications}
  \bibinfo{volume}{184} (\bibinfo{year}{2013}) \bibinfo{pages}{1234--1240}.
\bibitem[{Dalibard et~al.(1992)Dalibard, Castin, and Molmer}]{Dalibard:1992zz}
\bibinfo{author}{J.~Dalibard}, \bibinfo{author}{Y.~Castin},
  \bibinfo{author}{K.~Molmer},
\newblock \bibinfo{title}{{Wave-function approach to dissipative processes in
  quantum optics}},
\newblock \bibinfo{journal}{Phys. Rev. Lett.} \bibinfo{volume}{68}
  (\bibinfo{year}{1992}) \bibinfo{pages}{580--583}.
\bibitem[{M{\o}lmer et~al.(1993)M{\o}lmer, Castin, and Dalibard}]{Molmer:93}
\bibinfo{author}{K.~M{\o}lmer}, \bibinfo{author}{Y.~Castin},
  \bibinfo{author}{J.~Dalibard},
\newblock \bibinfo{title}{Monte carlo wave-function method in quantum optics},
\newblock \bibinfo{journal}{J. Opt. Soc. Am. B} \bibinfo{volume}{10}
  (\bibinfo{year}{1993}) \bibinfo{pages}{524--538}.
\bibitem[{Plenio and Knight(1998)}]{Plenio:1997ep}
\bibinfo{author}{M.~B. Plenio}, \bibinfo{author}{P.~L. Knight},
\newblock \bibinfo{title}{{The Quantum jump approach to dissipative dynamics in
  quantum optics}},
\newblock \bibinfo{journal}{Rev. Mod. Phys.} \bibinfo{volume}{70}
  (\bibinfo{year}{1998}) \bibinfo{pages}{101--144}.
\bibitem[{Carmichael(1999)}]{carmichael1999statistical}
\bibinfo{author}{H.~Carmichael}, \bibinfo{title}{Statistical methods in quantum
  optics : master equations and fokker-planck equations},
  \bibinfo{publisher}{Springer}, \bibinfo{address}{New York},
  \bibinfo{year}{1999}.
\bibitem[{Weiss(1993)}]{weissbook}
\bibinfo{author}{U.~Weiss}, \bibinfo{title}{Quantum Dissipative Systems},
  \bibinfo{publisher}{WORLD SCIENTIFIC}, \bibinfo{year}{1993}.
\bibitem[{Daley(2014)}]{Daley:2014fha}
\bibinfo{author}{A.~J. Daley},
\newblock \bibinfo{title}{{Quantum trajectories and open many-body quantum
  systems}},
\newblock \bibinfo{journal}{Adv. Phys.} \bibinfo{volume}{63}
  (\bibinfo{year}{2014}) \bibinfo{pages}{77--149}.
\bibitem[{Brambilla et~al.(2021)Brambilla, Escobedo, Strickland, Vairo,
  Vander~Griend, and Weber}]{Brambilla:2020qwo}
\bibinfo{author}{N.~Brambilla}, \bibinfo{author}{M.~A. Escobedo},
  \bibinfo{author}{M.~Strickland}, \bibinfo{author}{A.~Vairo},
  \bibinfo{author}{P.~Vander~Griend}, \bibinfo{author}{J.~H. Weber},
\newblock \bibinfo{title}{{Bottomonium suppression in an open quantum system
  using the quantum trajectories method}},
\newblock \bibinfo{journal}{JHEP} \bibinfo{volume}{05} (\bibinfo{year}{2021})
  \bibinfo{pages}{136}.
\bibitem[{Boyd et~al.(2019)Boyd, Cook, Islam, and Strickland}]{Boyd:2019arx}
\bibinfo{author}{J.~Boyd}, \bibinfo{author}{T.~Cook},
  \bibinfo{author}{A.~Islam}, \bibinfo{author}{M.~Strickland},
\newblock \bibinfo{title}{{Heavy quarkonium suppression beyond the adiabatic
  limit}},
\newblock \bibinfo{journal}{Phys. Rev.} \bibinfo{volume}{D100}
  (\bibinfo{year}{2019}) \bibinfo{pages}{076019}.
\bibitem[{Islam and Strickland(2020)}]{Islam:2020gdv}
\bibinfo{author}{A.~Islam}, \bibinfo{author}{M.~Strickland},
\newblock \bibinfo{title}{{Bottomonium suppression and elliptic flow from
  real-time quantum evolution}},
\newblock \bibinfo{journal}{Phys. Lett.} \bibinfo{volume}{B811}
  (\bibinfo{year}{2020}) \bibinfo{pages}{135949}.
\bibitem[{Islam and Strickland(2021)}]{Islam:2020bnp}
\bibinfo{author}{A.~Islam}, \bibinfo{author}{M.~Strickland},
\newblock \bibinfo{title}{{Bottomonium suppression and elliptic flow using
  Heavy Quarkonium Quantum Dynamics}},
\newblock \bibinfo{journal}{JHEP} \bibinfo{volume}{03} (\bibinfo{year}{2021})
  \bibinfo{pages}{235}.
\bibitem[{{Strickland et al}(2021)}]{qtraj-download}
\bibinfo{author}{{Strickland et al}}, \bibinfo{title}{{QTraj 1.0 - Public
  Repository}},
  \bibinfo{howpublished}{\url{https://bitbucket.org/kentphysics/qtraj-fftw}},
  \bibinfo{year}{2021}.
\bibitem[{{Intel Corporation}(2021)}]{IntelOneAPI}
\bibinfo{author}{{Intel Corporation}}, \bibinfo{title}{{Intel Math Kernel
  Library (Part of oneAPI)}},
  \bibinfo{howpublished}{\url{https://software.intel.com/content/www/us/en/develop/tools/oneapi/base-toolkit/download.html}},
  \bibinfo{year}{2021}.
\bibitem[{{Armadillo contributors}(2020)}]{Armadillo}
\bibinfo{author}{{Armadillo contributors}}, \bibinfo{title}{{Armadillo: C++
  library for linear algebra and scientific computing}},
  \bibinfo{howpublished}{\url{http://arma.sourceforge.net/download.html}},
  \bibinfo{year}{2020}.
\bibitem[{Caswell and Lepage(1986)}]{Caswell:1985ui}
\bibinfo{author}{W.~E. Caswell}, \bibinfo{author}{G.~P. Lepage},
\newblock \bibinfo{title}{{Effective Lagrangians for Bound State Problems in
  QED, QCD, and Other Field Theories}},
\newblock \bibinfo{journal}{Phys. Lett. B} \bibinfo{volume}{167}
  (\bibinfo{year}{1986}) \bibinfo{pages}{437--442}.
\bibitem[{Bodwin et~al.(1995)Bodwin, Braaten, and Lepage}]{Bodwin:1994jh}
\bibinfo{author}{G.~T. Bodwin}, \bibinfo{author}{E.~Braaten},
  \bibinfo{author}{G.~P. Lepage},
\newblock \bibinfo{title}{{Rigorous QCD analysis of inclusive annihilation and
  production of heavy quarkonium}},
\newblock \bibinfo{journal}{Phys. Rev. D} \bibinfo{volume}{51}
  (\bibinfo{year}{1995}) \bibinfo{pages}{1125--1171}. \bibinfo{note}{[Erratum:
  Phys.Rev.D 55, 5853 (1997)]}.
\bibitem[{Pineda and Soto(1998)}]{Pineda:1997bj}
\bibinfo{author}{A.~Pineda}, \bibinfo{author}{J.~Soto},
\newblock \bibinfo{title}{{Effective field theory for ultrasoft momenta in
  NRQCD and NRQED}},
\newblock \bibinfo{journal}{Nucl. Phys. B Proc. Suppl.} \bibinfo{volume}{64}
  (\bibinfo{year}{1998}) \bibinfo{pages}{428--432}.
\bibitem[{Brambilla et~al.(2000)Brambilla, Pineda, Soto, and
  Vairo}]{Brambilla:1999xf}
\bibinfo{author}{N.~Brambilla}, \bibinfo{author}{A.~Pineda},
  \bibinfo{author}{J.~Soto}, \bibinfo{author}{A.~Vairo},
\newblock \bibinfo{title}{{Potential NRQCD: An Effective theory for heavy
  quarkonium}},
\newblock \bibinfo{journal}{Nucl. Phys. B} \bibinfo{volume}{566}
  (\bibinfo{year}{2000}) \bibinfo{pages}{275}.
\bibitem[{Dumitru et~al.(2009)Dumitru, Guo, Mocsy, and
  Strickland}]{Dumitru:2009ni}
\bibinfo{author}{A.~Dumitru}, \bibinfo{author}{Y.~Guo},
  \bibinfo{author}{A.~Mocsy}, \bibinfo{author}{M.~Strickland},
\newblock \bibinfo{title}{{Quarkonium states in an anisotropic \protect{QCD}
  plasma}},
\newblock \bibinfo{journal}{Phys.Rev.} \bibinfo{volume}{D79}
  (\bibinfo{year}{2009}) \bibinfo{pages}{054019}.
\bibitem[{Strickland and Bazow(2012)}]{Strickland:2011aa}
\bibinfo{author}{M.~Strickland}, \bibinfo{author}{D.~Bazow},
\newblock \bibinfo{title}{{Thermal Bottomonium Suppression at RHIC and LHC}},
\newblock \bibinfo{journal}{Nucl. Phys.} \bibinfo{volume}{A879}
  (\bibinfo{year}{2012}) \bibinfo{pages}{25--58}.
\bibitem[{Breuer and Petruccione(2002)}]{Breuer:2002pc}
\bibinfo{author}{H.~Breuer}, \bibinfo{author}{F.~Petruccione},
  \bibinfo{title}{{The theory of open quantum systems}}, \bibinfo{year}{2002}.
\bibitem[{Gisin and Percival(1993)}]{Gisin:1992xc}
\bibinfo{author}{N.~Gisin}, \bibinfo{author}{I.~C. Percival},
\newblock \bibinfo{title}{{Quantum state diffusion, localization and quantum
  dispersion entropy}},
\newblock \bibinfo{journal}{J. Phys. A} \bibinfo{volume}{26}
  (\bibinfo{year}{1993}) \bibinfo{pages}{2233--2244}.
\bibitem[{Miura et~al.(2020)Miura, Akamatsu, Asakawa, and
  Rothkopf}]{Miura:2019ssi}
\bibinfo{author}{T.~Miura}, \bibinfo{author}{Y.~Akamatsu},
  \bibinfo{author}{M.~Asakawa}, \bibinfo{author}{A.~Rothkopf},
\newblock \bibinfo{title}{{Quantum Brownian motion of a heavy quark pair in the
  quark-gluon plasma}},
\newblock \bibinfo{journal}{Phys. Rev. D} \bibinfo{volume}{101}
  (\bibinfo{year}{2020}) \bibinfo{pages}{034011}.
\bibitem[{Sharma and Tiwari(2020)}]{Sharma:2019xum}
\bibinfo{author}{R.~Sharma}, \bibinfo{author}{A.~Tiwari},
\newblock \bibinfo{title}{{Quantum evolution of quarkonia with correlated and
  uncorrelated noise}},
\newblock \bibinfo{journal}{Phys. Rev. D} \bibinfo{volume}{101}
  (\bibinfo{year}{2020}) \bibinfo{pages}{074004}.
\bibitem[{Fornberg and Whitham(1978)}]{Fornberg:1978}
\bibinfo{author}{B.~Fornberg}, \bibinfo{author}{G.~Whitham},
\newblock \bibinfo{title}{A numerical and theoretical study of certain
  nonlinear wave phenomena},
\newblock \bibinfo{journal}{Phil. Trans. Roy. Soc.} \bibinfo{volume}{289}
  (\bibinfo{year}{1978}) \bibinfo{pages}{373}.
\bibitem[{Taha and Ablowitz(1984)}]{TAHA1984203}
\bibinfo{author}{T.~R. Taha}, \bibinfo{author}{M.~I. Ablowitz},
\newblock \bibinfo{title}{Analytical and numerical aspects of certain nonlinear
  evolution equations. ii. numerical, nonlinear schr\"odinger equation},
\newblock \bibinfo{journal}{Journal of Computational Physics}
  \bibinfo{volume}{55} (\bibinfo{year}{1984}) \bibinfo{pages}{203--230}.
\bibitem[{{FFTW contributors}(2020)}]{FFTW}
\bibinfo{author}{{FFTW contributors}}, \bibinfo{title}{{FFTW: A C subroutine
  library for computing the discrete Fourier transform}},
  \bibinfo{howpublished}{\url{http://www.fftw.org/}}, \bibinfo{year}{2020}.
\bibitem[{Casalderrey-Solana and Teaney(2006)}]{CasalderreySolana:2006rq}
\bibinfo{author}{J.~Casalderrey-Solana}, \bibinfo{author}{D.~Teaney},
\newblock \bibinfo{title}{{Heavy quark diffusion in strongly coupled N=4
  Yang-Mills}},
\newblock \bibinfo{journal}{Phys. Rev. D} \bibinfo{volume}{74}
  (\bibinfo{year}{2006}) \bibinfo{pages}{085012}.
\bibitem[{Caron-Huot and Moore(2008)}]{CaronHuot:2007gq}
\bibinfo{author}{S.~Caron-Huot}, \bibinfo{author}{G.~D. Moore},
\newblock \bibinfo{title}{{Heavy quark diffusion in perturbative \protect{QCD}
  at next-to- leading order}},
\newblock \bibinfo{journal}{Phys. Rev. Lett.} \bibinfo{volume}{100}
  (\bibinfo{year}{2008}) \bibinfo{pages}{052301}.
\bibitem[{Brambilla et~al.(2019)Brambilla, Escobedo, Vairo, and
  Vander~Griend}]{Brambilla:2019tpt}
\bibinfo{author}{N.~Brambilla}, \bibinfo{author}{M.~A. Escobedo},
  \bibinfo{author}{A.~Vairo}, \bibinfo{author}{P.~Vander~Griend},
\newblock \bibinfo{title}{{Transport coefficients from in medium quarkonium
  dynamics}},
\newblock \bibinfo{journal}{Phys. Rev. D} \bibinfo{volume}{100}
  (\bibinfo{year}{2019}) \bibinfo{pages}{054025}.
\bibitem[{Brambilla et~al.(2020)Brambilla, Leino, Petreczky, and
  Vairo}]{Brambilla:2020siz}
\bibinfo{author}{N.~Brambilla}, \bibinfo{author}{V.~Leino},
  \bibinfo{author}{P.~Petreczky}, \bibinfo{author}{A.~Vairo},
\newblock \bibinfo{title}{{Lattice QCD constraints on the heavy quark diffusion
  coefficient}},
\newblock \bibinfo{journal}{Phys. Rev. D} \bibinfo{volume}{102}
  (\bibinfo{year}{2020}) \bibinfo{pages}{074503}.
\bibitem[{Inc.(2020)}]{Mathematica}
\bibinfo{author}{W.~R. Inc.}, \bibinfo{title}{Mathematica, {V}ersion 12.1},
  \bibinfo{year}{2020}. \URLprefix \url{https://www.wolfram.com/mathematica}.
\bibitem[{Pordes et~al.(2007)Pordes, Petravick, Kramer, Olson, Livny, Roy,
  Avery, Blackburn, Wenaus, W{\"u}rthwein, Foster, Gardner, Wilde, Blatecky,
  McGee, and Quick}]{osg07}
\bibinfo{author}{R.~Pordes}, \bibinfo{author}{D.~Petravick},
  \bibinfo{author}{B.~Kramer}, \bibinfo{author}{D.~Olson},
  \bibinfo{author}{M.~Livny}, \bibinfo{author}{A.~Roy},
  \bibinfo{author}{P.~Avery}, \bibinfo{author}{K.~Blackburn},
  \bibinfo{author}{T.~Wenaus}, \bibinfo{author}{F.~W{\"u}rthwein},
  \bibinfo{author}{I.~Foster}, \bibinfo{author}{R.~Gardner},
  \bibinfo{author}{M.~Wilde}, \bibinfo{author}{A.~Blatecky},
  \bibinfo{author}{J.~McGee}, \bibinfo{author}{R.~Quick},
\newblock \bibinfo{title}{The open science grid},
\newblock in: \bibinfo{booktitle}{J. Phys. Conf. Ser.},
  volume~\bibinfo{volume}{78} of \textit{\bibinfo{series}{78}},
  \bibinfo{year}{2007}, p. \bibinfo{pages}{012057}.
  \DOIprefix\doi{10.1088/1742-6596/78/1/012057}.
\bibitem[{Sfiligoi et~al.(2009)Sfiligoi, Bradley, Holzman, Mhashilkar, Padhi,
  and Wurthwein}]{osg09}
\bibinfo{author}{I.~Sfiligoi}, \bibinfo{author}{D.~C. Bradley},
  \bibinfo{author}{B.~Holzman}, \bibinfo{author}{P.~Mhashilkar},
  \bibinfo{author}{S.~Padhi}, \bibinfo{author}{F.~Wurthwein},
\newblock \bibinfo{title}{The pilot way to grid resources using glideinwms},
\newblock in: \bibinfo{booktitle}{2009 WRI World Congress on Computer Science
  and Information Engineering}, volume~\bibinfo{volume}{2} of
  \textit{\bibinfo{series}{2}}, \bibinfo{year}{2009}, pp.
  \bibinfo{pages}{428--432}. \DOIprefix\doi{10.1109/CSIE.2009.950}.
\bibitem[{{Strickland et al}(2021)}]{kent-code-library}
\bibinfo{author}{{Strickland et al}}, \bibinfo{title}{{Sample trajectories for
  use in QTraj 1.0}},
  \bibinfo{howpublished}{\url{http://personal.kent.edu/~mstrick6/code}},
  \bibinfo{year}{2021}.
\bibitem[{Brambilla et~al.( RTG)Brambilla, Escobedo, Strickland, Vairo,
  Vander~Griend, and Weber}]{munichforth}
\bibinfo{author}{N.~Brambilla}, \bibinfo{author}{M.~A. Escobedo},
  \bibinfo{author}{M.~Strickland}, \bibinfo{author}{A.~Vairo},
  \bibinfo{author}{P.~Vander~Griend}, \bibinfo{author}{J.~H. Weber},
  \bibinfo{title}{{Bottomonium production in heavy-ion collisions using quantum
  trajectories: Differential observables and momentum anisotropy}},
  \bibinfo{year}{TUM-EFT 147/21, HU-EP-21/18-RTG}.
\bibitem[{{Free Software Foundation}(2019)}]{GSL}
\bibinfo{author}{{Free Software Foundation}}, \bibinfo{title}{{GNU Scientific
  Library}}, \bibinfo{howpublished}{\url{https://www.gnu.org/software/gsl/}},
  \bibinfo{year}{2019}.
\bibitem[{{Google, Inc.}(2020)}]{Googletest}
\bibinfo{author}{{Google, Inc.}}, \bibinfo{title}{{GoogleTest}},
  \bibinfo{howpublished}{\url{https://github.com/google/googletest}},
  \bibinfo{year}{2020}.
\bibitem[{{Kitware, Inc. and contributors}(2020)}]{cmake}
\bibinfo{author}{{Kitware, Inc. and contributors}}, \bibinfo{title}{{Cmake}},
  \bibinfo{howpublished}{\url{https://cmake.org/download/}},
  \bibinfo{year}{2020}.
\bibitem[{{Homebrew contributors}(2020)}]{homebrew}
\bibinfo{author}{{Homebrew contributors}}, \bibinfo{title}{{Homebrew}},
  \bibinfo{howpublished}{\url{https://brew.sh}}, \bibinfo{year}{2020}.
\bibitem[{{Veridian Information Solutions}(2000)}]{openpbs}
\bibinfo{author}{{Veridian Information Solutions}}, \bibinfo{title}{{OpenPBS}},
  \bibinfo{howpublished}{\url{https://www.openpbs.org/}}, \bibinfo{year}{2000}.
\bibitem[{{HTCondor team}(2021)}]{condor}
\bibinfo{author}{{HTCondor team}}, \bibinfo{title}{{HTCondor}},
  \bibinfo{howpublished}{\url{https://research.cs.wisc.edu/htcondor/}},
  \bibinfo{year}{2021}.
\bibitem[{Nopoush et~al.(2014)Nopoush, Ryblewski, and
  Strickland}]{Nopoush:2014pfa}
\bibinfo{author}{M.~Nopoush}, \bibinfo{author}{R.~Ryblewski},
  \bibinfo{author}{M.~Strickland},
\newblock \bibinfo{title}{{Bulk viscous evolution within anisotropic
  hydrodynamics}},
\newblock \bibinfo{journal}{Phys.Rev.} \bibinfo{volume}{C90}
  (\bibinfo{year}{2014}) \bibinfo{pages}{014908}.
\bibitem[{Bazavov et~al.(2016)Bazavov, Brambilla, Ding, Petreczky, Schadler,
  Vairo, and Weber}]{Bazavov:2016uvm}
\bibinfo{author}{A.~Bazavov}, \bibinfo{author}{N.~Brambilla},
  \bibinfo{author}{H.~T. Ding}, \bibinfo{author}{P.~Petreczky},
  \bibinfo{author}{H.~P. Schadler}, \bibinfo{author}{A.~Vairo},
  \bibinfo{author}{J.~H. Weber},
\newblock \bibinfo{title}{{Polyakov loop in 2+1 flavor QCD from low to high
  temperatures}},
\newblock \bibinfo{journal}{Phys. Rev. D} \bibinfo{volume}{93}
  (\bibinfo{year}{2016}) \bibinfo{pages}{114502}.
\bibitem[{Bazavov et~al.(2019)}]{Bazavov:2018mes}
\bibinfo{author}{A.~Bazavov}, et~al. (\bibinfo{collaboration}{HotQCD}),
\newblock \bibinfo{title}{{Chiral crossover in QCD at zero and non-zero
  chemical potentials}},
\newblock \bibinfo{journal}{Phys. Lett. B} \bibinfo{volume}{795}
  (\bibinfo{year}{2019}) \bibinfo{pages}{15--21}.
\bibitem[{Borsanyi et~al.(2020)Borsanyi, Fodor, Guenther, Kara, Katz, Parotto,
  Pasztor, Ratti, and Szabo}]{Borsanyi:2020fev}
\bibinfo{author}{S.~Borsanyi}, \bibinfo{author}{Z.~Fodor},
  \bibinfo{author}{J.~N. Guenther}, \bibinfo{author}{R.~Kara},
  \bibinfo{author}{S.~D. Katz}, \bibinfo{author}{P.~Parotto},
  \bibinfo{author}{A.~Pasztor}, \bibinfo{author}{C.~Ratti},
  \bibinfo{author}{K.~K. Szabo},
\newblock \bibinfo{title}{{QCD Crossover at Finite Chemical Potential from
  Lattice Simulations}},
\newblock \bibinfo{journal}{Phys. Rev. Lett.} \bibinfo{volume}{125}
  (\bibinfo{year}{2020}) \bibinfo{pages}{052001}.
\bibitem[{McNeile et~al.(2010)McNeile, Davies, Follana, Hornbostel, and
  Lepage}]{McNeile:2010ji}
\bibinfo{author}{C.~McNeile}, \bibinfo{author}{C.~T.~H. Davies},
  \bibinfo{author}{E.~Follana}, \bibinfo{author}{K.~Hornbostel},
  \bibinfo{author}{G.~P. Lepage},
\newblock \bibinfo{title}{{High-Precision c and b Masses, and QCD Coupling from
  Current-Current Correlators in Lattice and Continuum QCD}},
\newblock \bibinfo{journal}{Phys. Rev.} \bibinfo{volume}{D82}
  (\bibinfo{year}{2010}) \bibinfo{pages}{034512}.
\bibitem[{Strickland(2018)}]{Strickland:2018ayk}
\bibinfo{author}{M.~Strickland},
\newblock \bibinfo{title}{{The non-equilibrium attractor for kinetic theory in
  relaxation time approximation}},
\newblock \bibinfo{journal}{JHEP} \bibinfo{volume}{12} (\bibinfo{year}{2018})
  \bibinfo{pages}{128}.
\bibitem[{Islam and Strickland(ming)}]{islamforth}
\bibinfo{author}{A.~Islam}, \bibinfo{author}{M.~Strickland},
  \bibinfo{title}{{Including momentum-space anisotropy in real-time quantum
  mechanical calculations of quarkonium suppression}},
  \bibinfo{year}{forthcoming}.
\bibitem[{Romatschke and Strickland(2003)}]{Romatschke:2003ms}
\bibinfo{author}{P.~Romatschke}, \bibinfo{author}{M.~Strickland},
\newblock \bibinfo{title}{Collective modes of an anisotropic quark gluon
  plasma},
\newblock \bibinfo{journal}{Phys. Rev.} \bibinfo{volume}{D68}
  (\bibinfo{year}{2003}) \bibinfo{pages}{036004}.
\bibitem[{Martinez et~al.(2012)Martinez, Ryblewski, and
  Strickland}]{Martinez:2012tu}
\bibinfo{author}{M.~Martinez}, \bibinfo{author}{R.~Ryblewski},
  \bibinfo{author}{M.~Strickland},
\newblock \bibinfo{title}{{Boost-Invariant (2+1)-dimensional Anisotropic
  Hydrodynamics}},
\newblock \bibinfo{journal}{Phys. Rev.} \bibinfo{volume}{C85}
  (\bibinfo{year}{2012}) \bibinfo{pages}{064913}.

\end{thebibliography}

\end{document}